\documentclass[letterpaper,openright,11pt]{article}

\usepackage[english]{babel}
\usepackage[utf8]{inputenc}
\usepackage{sidecap,mathtools,leftindex}
\usepackage{hyperref,cite}
\usepackage{mathrsfs,ragged2e}
\usepackage{authblk,blindtext}
\usepackage{booktabs,multirow} 
\usepackage{caption,subcaption}
\usepackage[usenames]{xcolor}
\usepackage{amsmath,amssymb,latexsym}
\usepackage{soul}
\usepackage{physics,bm}
\usepackage{color,cancel,graphicx,bbm,colortbl}
\usepackage[titletoc,toc,title]{appendix}
\usepackage[left=2.0cm,top=2.5cm,right=2.0cm,bottom=2.5cm]{geometry}

\graphicspath{ {FigurasPaperCF/} }

\providecommand{\keywords}[1]{\small \textbf{\textit{Keywords---}} #1}

\title{Theoretical analysis of the effect of the kinetic energy operator on double heterostructure spectra}
\author[1]{R. Valencia-Torres} 
\author[2]{J. García-Ravelo}
\author[1]{E. Choreño-Ortiz}
\author[2]{J. Avendaño}
\affil[1]{\small Centro de Investigación y de Estudios Avanzados del IPN, Departamento de Física, Ciudad de México, 07360, México.}  
\affil[2]{\small Instituto Politécnico Nacional, ESFM, Departamento de Física, Ciudad de México, 07738, México.}
\date{}

\begin{document}
\maketitle

\begin{abstract}
We report an analytical and numerical study of the effect of kinetic energy operator (KEO) ambiguity on the energy spectra of double heterostructures when the mass of the charge carriers, subjected to a potential, depends on position. 
The spectra are calculated using two complementary techniques. 
In the first case, which is not always possible, the energy spectrum satisfies a transcendental equation, which is obtained through an analytical process. 
While in the second, which can be used practically in any double heterostructure (DH), the energy spectra correspond to the poles of the reflection coefficient of the heterostructure. 
The spectra thus obtained complement those already reported in a previous reference, since we now include the spectra of another KEO. 
Finally, we show how these methods allow us to critically analyze some statements that have been made about the relevance of some kinetic energy operators in some simpler heterostructures.
\end{abstract}


\section{Introduction} \label{Intro}
In condensed matter physics, the study of heterostructures requires the analysis of low-energy excitations of charge carriers. 
These excitations are described by an effective Schrödinger equation, where the KEO is an ambiguous quantity. 
This ambiguity arises because the mass operator $m=M_{0} m(z)$, no longer commutes with the momentum operator $\hat{p}=-i\hbar d/dz$. 
As a result, the correct ordering of these operators in the kinetic energy term is not uniquely defined, leading to the von Roos KEO \cite{PhysRevB.27.7547}

\begin{eqnarray}
T_{\textrm{vR}}(\alpha,\gamma)=\frac{1}{4}  (m^{\alpha}\hat{p}m^{\beta}\hat{p}m^{\gamma}+m^{\gamma}\hat{p}m^{\beta}\hat{p}m^{\alpha}),
\label{kineticop}
\end{eqnarray}
without a defined ordering, because to guarantee its hermiticity,  $\alpha+\beta+\gamma =-1$.
This problem of the ambiguity of the von Roos KEO has been widely discussed in Refs. \cite{PhysRevB.39.12783,DESOUZADUTRA200025,Valencia,PhysRevB.30.678,PhysRevB.35.8074,GTEinevoll1988,HAGSTON1994199,PhysRevB.51.17624,PhysRevB.53.16504,refId0,PhysRevLett.80.3823,PhysRevB.60.14269,LEWYANVOON2002269,PhysRevB.65.165328,HarrisonWalter,GADELLA2007265,GADELLA20091310,KURASOV1996297,PhysRevB.55.1326,MustafaOAR}, from different perspectives. 
In this work, we address some approaches that, despite having been replicated in some references, must be treated critically.
For example, Ref. \cite{PhysRevB.39.12783} concludes that energy eigenvalues of a particle jumping in mass within a finite one-dimensional box, diverge “unless $\alpha=\gamma$ and $\beta=-1$ is the only universal set of operator ordering parameters” with unique results.
On the other hand, in the Ref. \cite{DESOUZADUTRA200025}, some KEOs have been discarded.
The main idea to do this is ``to suppose that, once one have found the ordering without ambiguity for a given potential or class of potentials, that ordering should be extended to remaining physical potentials" and it warns that for an unconfined particle with position-depend mass its energy spectrum is ``not well defined" for some of the orderings where $\alpha \neq \gamma$ and $\alpha=\gamma$.\\ 
For the purposes of this work, we distinguish between two types of heterostructures. 
A graded heterostructure (GH) is characterized by potential $V(z)$ and mass $m(z)$ distributions that vary smoothly and continuously across the entire structure.
In contrast, a double heterostructure (DH) consists of three regions: in the intermediate region ($z_{0}<z<z_{1}$)  both the mass $m_{\textrm{in}}(z)$ and the potential $V_{\textrm{in}}(z)$ are, while in the outer the mass and potential remain constant, that is
\begin{eqnarray}
V(z)=
\begin{cases}
V_{0} \\
V_{\textrm{in}}(z) \\
V_{2}
\end{cases}, \quad
m(z)=
\begin{cases}
m_{0}     &; \quad z\leq z_{0}, \\
m_{\textrm{in}}(z) &; \quad z_{0}< z < z_{1}, \label{hetmodel}\\
m_{2}     &; \quad z_{1}\leq z.
\end{cases}
\label{modelmass}
\end{eqnarray}

In Ref.  \cite{Valencia}, following two different but complementary strategies, the \textit{effect of KEO ambiguity on double heterostructures spectra} had been estimated for KEOs with $\alpha=\gamma$.
In both procedures, the corresponding boundary conditions for the wave function of the charge carriers at points where there is a finite abrupt change in mass must be taken into account. 
That is, at the  points where there is an abrupt finite change of mass (and / or potential) defined by the Heaviside function, $H$, as
\begin{eqnarray}
m(z) = m_{k}H(z_{k} - z) + m_{k+1}H(z - z_{k}),
\label{masaf}
\end{eqnarray}
and at the points $z_{0,1}$ of the DH \eqref{modelmass} when $m_{\textrm{in}}(z_{0,1})\neq m_{0,2}$.

In this Ref. \cite{Valencia}, only the ambiguities $\alpha=\gamma=0$ of BenDaniel-Duke (BD-D) \cite{PhysRev.152.683} and $\alpha=\gamma=-1/2$ of Zhu-Kroemer (Z-K) \cite{PhysRevB.27.3519}  were considered without taking into account choices such as $\beta=\gamma=-1/2$ of Li-Kuhn (L-K) \cite{PhysRevB.47.12760} where $\alpha \neq \gamma$.
Other examples of orderings where $\alpha \neq \gamma$ are given by Gora and Williams (G-W) \cite{PhysRev.177.1179} and the ordering that gives rise to the  operator \cite{lima}
\begin{equation}
T_{L} =\frac{1}{6} \left(m^{-1} \hat{p}^{2} + \hat{p} m^{-1} \hat{p} + \hat{p}^2 m^{-1} \right),
\label{NewOpn} 
\end{equation}
which has the ambiguity defined by conditions $\alpha+\gamma= -\frac{2}{3}$, $\alpha \gamma=0$. When $m(z)$ and its derivatives are continuous, $T_{L}$ is a particular case of von Roos KEO with $\alpha \neq \gamma$,
see Eqs. \eqref{eq:17} and \eqref{eq:18} in Appendix \ref{AppenConFronA}.
For this KEO, the boundary conditions for the wave function $\psi(z)$ of charge carriers at points $z_{k}$ within the heterostructure, are \cite{lima}
\begin{eqnarray}
\psi(z_{k}^{-})=\frac{2m_{k}+m_{k+1}}{m_{k}+2m_{k+1}}\psi(z_{k}^{+}); \quad
\psi'(z_{k}^{-})=\frac{5m_{k}+m_{k+1}}{m_{k}+5m_{k+1}}\psi'(z_{k}^{+}),
\label{bcondNew}
\end{eqnarray}
which are obtained again through the arguments in Appendix \ref{AppenConFronA}.

Despite the warning in Ref. \cite{PhysRevB.39.12783}, the question arises: to which energy spectrum does a forbidden ordering lead when using any of the strategies of Ref. \cite{Valencia} for the DHs in this work?.
Even more so, if cases with alpha different from gamma don’t guarantee the conservation of probability flux at an interface.

This paper is organized as follows.
Section \ref{SchPDM} briefly analyzes the ambiguity that leads to the KEO defined by Eq. \eqref{NewOpn} and poses the Schrödinger equation for the DH model \eqref{modelmass}.
Section \ref{RefCoef} generalizes the strategies from Ref. \cite{Valencia}, to now include the DHs energy spectra with orderings $\alpha \neq \gamma$.
Section \ref{DobHet}, presents new energy spectra for the DHs previously studied in Ref. \cite{Valencia}, focusing on cases with operator \eqref{NewOpn} and corresponding boundary conditions.
This section also includes two DHs, constructed from the GH distributions of Ref. \cite{DESOUZADUTRA200025}. 
Additionally, we make some observations about this reference.
Finally, Section \ref{concl}, summarizes our main findindings based on the energy spectra obtained with three different KEO ambiguities.

\section{Schrödinger equation for position-dependent effective masses} \label{SchPDM}
Before formulating the Schrödinger equation for the DH \eqref{modelmass}, let us observe the connection existing between the operators given by Eqs. \eqref{kineticop} and \eqref{NewOpn}, rewriting them in a comparable form\footnote{In the following we adopt units $\hbar^{2} (2M_{0})^{-1}=1$ and $m'(z)$ denotes the derivative of $m(z)$ with respect to $z$.}:
\begin{eqnarray}
T_{\textrm{vR}}(\alpha,\gamma)=-\frac{d}{dz}\frac{1}{m(z)}\frac{d}{dz} +V_{\alpha \gamma}(z); \quad
V_{\alpha \gamma}(z)= -\frac{1}{2}\left( \nu \frac{m''(z)}{m(z)^{2}}-\eta\frac{m'(z)^{2}}{m(z)^{3}}\right),
\label{KEOvonRoos}
\end{eqnarray}
$\nu=\alpha+\gamma, \quad \eta=2(\alpha+\gamma+\alpha\gamma)$, and
\begin{eqnarray}
T_{\textrm{L}} = -\frac{d}{dz}\frac{1}{m(z)}\frac{d}{dz}+V_{L}(z); \quad
V_{L}(z)=\frac{1}{3} \left(\frac{m''(z)}{m(z)^{2}}-2\frac{m'(z)^{2}}{m(z)^{3}}\right),
\label{KEOLima}
\end{eqnarray}
respectively. We observe that the operator $T_{\textrm{L}}$ is actually a particular case of the von Roos operator $T_{\textrm{vR}}(\alpha,\gamma)$ when $m(z)$ and its derivatives are continuos functions, provided that the parameters 
\begin{eqnarray}
\nu=\alpha+\gamma= -\frac{2}{3}; \quad \eta=2(\alpha+\gamma+\alpha\gamma)=-\frac{4}{3}.
\label{etanuL}
\end{eqnarray}
That is, $\alpha$ and $\gamma$ must satisfy $\alpha+\gamma= -\frac{2}{3}$ and $\alpha\gamma=0$ (i.e. $\alpha \neq \gamma$).
This conclusion is significant because it contrasts with the statement in Ref. \cite{lima} where it was claimed that the $T_{L}$ operator does not belong to the von Roos family.

Thus, the Schr\"odinger equation 
\begin{eqnarray}
\left(- \frac{d}{dz}\frac{1}{m(z)}\frac{d}{dz}+V_{\alpha \gamma}(z) + V(z) - E\right) \psi= 0,
\label{shoeqn}
\end{eqnarray}
allow us to obtain the quantum states of the charge carrier of mass $m(z)$.
As far as the energy spectrum is concerned, $E$, the boundary conditions of $\psi$ are relevant.
At the points $z_{k}$, where the mass satisfies \eqref{masaf}, the respective boundary conditions must be considered.
In Ref. \cite{Valencia}, where the von Roos KEO was considered with $\alpha=\gamma$, the boundary conditions  for the wave function $\psi$ and its derivative at these points  are given by \cite{lima}
\begin{eqnarray}
\psi(z_{k}^{-})=\frac{m_{k+1}^{\alpha+1}m_{k}^{\alpha+\beta+1}+m_{k}}{m_{k}^{\alpha+1}m_{k+1}^{\alpha+\beta+1}+m_{k+1}}\psi(z_{k}^{+}); \quad
\psi'(z_{k}^{-})=\frac{m_{k}^{\alpha+1}m_{k+1}^{\alpha+\beta+1}+m_{k}}{m_{k+1}^{\alpha+1}m_{k}^{\alpha+\beta+1}+m_{k+1}}\psi'(z_{k}^{+}).
\label{bcondvonR}
\end{eqnarray}
On the other hand, the corresponding boundary conditions used for the KEO \eqref{KEOLima} are the Eqs. \eqref{bcondNew}.
In fact, Eqs. \eqref{bcondNew} and \eqref{bcondvonR} explicitly show that each ordering of KEO, provides particular boundary conditions to the wave function at points $z_{k}$, which, as we have said, causes an effect on the energy spectrum $E$. 

An alternative approach\footnote{A variation of the method is outlined in Appendix \ref{AppendB}, applied to the DHs specified in Tables \ref{TFigT8} and \ref{FigT9}.} to solving Eq. \eqref{shoeqn} involves transforming it into an associated Schrödinger equation with constant (unit) mass. 
This is achieved by defining
\begin{eqnarray}
\psi(z)=m(z)^{1/4}\phi, \quad \rho=\int \sqrt{m(z)}dz,
\label{transf}
\end{eqnarray}

where the new variable $\rho$ provided that the function $m(z)$ allows for the inverse transformation $z(\rho)$. The function $\phi$ then the isospectral Schr\"odinger equation with constant mass
\begin{eqnarray}
-\frac{d^2\phi}{d\rho^2}+(\widetilde{V}(\rho)-E)\phi=0,
\label{SchomassC}
\end{eqnarray}
and effective potential 
\begin{eqnarray}
\widetilde{V}(\rho)=V(z) + \frac{1}{2} \left(\eta+\frac{7}{8}\right)\frac{m'(z)^{2}}{m(z)^{3}}-\frac{1}{2} \left(\nu + \frac{1}{2}\right)\frac{m''(z)}{m(z)^{2}}.
\label{Vtilde}
\end{eqnarray}

If $V(z)$ and $m(z)$, are chosen appropriately, this transformed equation may correspond to a well-known or simpler problem, allowing the energy spectrum $E$ to be determined directly.
The original wave function $\psi$ can then be reconstructed from  $\phi$ and $m$ using the transformation \eqref{transf}. For smooth profiles, it is important to apply the `general' boundary conditions as we will see in Section \ref{AnalyMetho}.
This technique has already been applied in Ref. \cite{Valencia} for two DHs revisited in Secs. \ref{symdist} and \ref{SMorseDist}, and in Ref. \cite{DESOUZADUTRA200025} for two GHs to evaluate the validity of different KEOs. 
In this work, each of these two GHs, is embedded within a DH (Eq. \eqref{modelmass}), as described in Secs. \ref{SExpMass}, and \ref{SubSingMass}. 
Since that from our DH models can be reduced to the GH model, we critically analyze the procedure followed in Ref \cite{DESOUZADUTRA200025} to discard certain KEOs.

\section{Two strategies for calculating energy spectra in DHs} \label{RefCoef}
The DH model we have adopted \eqref{modelmass} can be seen from two complementary points of view. 
In the first, when the Eq. \eqref{KEOvonRoos} is taken into account, the Schrödinger equation \eqref{shoeqn}
\begin{eqnarray}
\left(- \frac{d}{dz}\frac{1}{m_{\textrm{in}}(z)}\frac{d}{dz}
-\frac{1}{2}\left( \nu \frac{m_{\textrm{in}}''(z)}{m_{\textrm{in}}(z)^{2}}-\eta\frac{m_{\textrm{in}}'(z)^{2}}{m_{\textrm{in}}(z)^{3}}\right) +
 V_{\textrm{in}}(z) - E\right) \psi_{\textrm{in}}= 0,
\label{shoeqn3}
\end{eqnarray}
must be resolved by the wave function of carriers. 
In this equation, the spatial distributions involved may or may not be continuous at the points $z_{k}$ of the heterojunction.
In practice, this equation can be solved analytically only for some potential and mass distributions.

In the second point of view, practically any pair of distributions $V(z)$ and $m(z)$ not necessarily confined, can be approximated as a succession of abrupt finite steps in potential and mass
\begin{eqnarray}
V(z)=
\begin{cases}
V_{0} \\
{V_{\textrm{in}}}_{1} \\
\cdot \\
{V_{\textrm{in}}}_{j} \\
\cdot \\
V_{2} \\
\end{cases}, \quad
m(z)=
\begin{cases}
m_{0}; \quad &z<z_{0}, \\
{m_{\textrm{in}}}_{1}; \quad &z_{0}\leq z<{z_{\textrm{in}}}_{1}, \\
\cdot \\
{m_{\textrm{in}}}_{j}; \quad &{z_{\textrm{in}}}_{j-1}\leq z<{z_{\textrm{in}}}_{j}, \\
\cdot \\
m_{2}; \quad &z_{1}\leq z. \\
\end{cases}
\label{potmassdis}
\end{eqnarray}
${z_{\textrm{in}}}_{j}$; $j=0,1,\dots,n$ (${z_{\textrm{in}}}_{0}=z_{0},  {z_{\textrm{in}}}_{n}=z_{1}$) represents a point  at the intermediate region, where there is a finite abrupt change of potential and mass. 
Since charge carriers with position-dependent effective mass $m(z)$ within the DH are subject to the potential $V(z)$, their probability amplitude must be a solution of the simple Schrödinger equation
\begin{eqnarray}
-\frac{1}{m_{\textrm{in}j}}\frac{d^2}{dz^2}\psi_{\textrm{in}j}+(V_{\textrm{in}j}-E)\psi_{\textrm{in}j}=0,
\label{SchrEqj}
\end{eqnarray}
in each interval $[{z_{\textrm{in}}}_{j-1}, {z_{\textrm{in}}}_{j}]$. 
Evidently, the probability amplitude $\psi_{\textrm{in}j}$ of each interval must be joined with the neighboring amplitudes $\psi_{\textrm{in}j-1}$ and $\psi_{\textrm{in}j+1}$, using the boundary conditions \eqref{bcondNew} and \eqref{bcondvonR} that we have available for this work.

\subsection{First strategy: `analytic' method} \label{AnalyMetho}
According to Ref. \cite{Valencia}, this is an analytical method to solve the Schrödinger equation \eqref{shoeqn3}. 
It can be used to verify the accuracy of the numerical methodology that will be presented in Section \ref{miltistepm}. 
We will achieve this by comparing the `analytical' energy spectrum of a charge carrier with the poles of the reflection coefficient of the DH, as was previously done in Ref. \cite{Valencia} for cases $\alpha=\gamma$.
In the inner region, the particle obeys Eq. \eqref{shoeqn3}, so the complete solution, in the case of bound states, can be written as 
\begin{eqnarray}
\psi(z)=
\begin{cases}
\psi_{0}(z)=R e^{\eta_{0}z} &;\quad z<z_{0}, \\
\psi_{\textrm{in}}(z)=P\psi^{1}_{\textrm{in}}(z)+ Q\psi^{2}_{\textrm{in}}(z) &; \quad z_{0}\leq z \leq z_{1},\\
\psi_{2}(z)=T e^{-\eta_{2}z}  &; \quad z_{1}<z,
\end{cases}
 \label{gensol}
\end{eqnarray}
where $\eta_{0,2}=\sqrt{m_{0,2}(V_{0,2}-E)}$. At points, $z_{0}$ and $z_{1}$ we must use boundary conditions (see. Ref. \cite{Valencia}) pending of the KEO we have chosen. Those conditions get us an equation system that we can write as
\begin{eqnarray}
\begin{cases}
R e^{\eta_{0}z_{0}}=P\psi^{1}_{\textrm{in}}(z_{0})+ Q\psi^{2}_{\textrm{in}}(z_{0})=-P\Phi_{11}-Q\Phi_{12},\\
T e^{-\eta_{2}z_{1}}=P\psi^{1}_{\textrm{in}}(z_{1})+ Q\psi^{2}_{\textrm{in}}(z_{1}) =-P\Phi_{21}-Q\Phi_{22}.
\end{cases}
\label{homsys1}
\end{eqnarray}
This homogeneous system can be written in matrix form as
\begin{eqnarray}
X \left(\begin{array}{c}P \\Q\end{array}\right):=
\left(\begin{array}{cc}
\psi^{1}_{\textrm{in}}(z_{0})+\Phi_{11} & \psi^{2}_{\textrm{in}}(z_{0})+\Phi_{12} \\
\psi^{1}_{\textrm{in}}(z_{1})+\Phi_{21} & \psi^{2}_{\textrm{in}}(z_{1})+\Phi_{22}
\end{array}\right)
 \left(\begin{array}{c}P \\Q\end{array}\right)=0.
 \label{matrixhom}
\end{eqnarray}  
In general, these coefficients (functions) $\Phi_{ji}$, rely on the KEO's choice. For the BD-D KEO we get
\begin{eqnarray}
\Phi_{1i}=-\frac{1}{\eta_{0}}\frac{m_{0}}{m_{\textrm{in}(z_{0})}}\left(\psi^{i}_{\textrm{in}}(z)\right)'|_{z=z_{0}}, \quad
\Phi_{2i}=\frac{1}{\eta_{2}}\frac{m_{2}}{m_{\textrm{in}(z_{1})}}\left(\psi^{i}_{\textrm{in}}(z)\right)'|_{z=z_{1}}; \quad i=1,2. 
\label{BDDconditions}
\end{eqnarray}
For the Z-K KEO
\begin{eqnarray}
\Phi_{1i}=
-\frac{\sqrt{m_{\textrm{in}(z_{0})}}}{\eta_{0}}
\left(\frac{\psi^{i}_{\textrm{in}}(z)}{\sqrt{m_{\textrm{in}(z)}}}\right)'|_{z=z_{0}}, \quad
\Phi_{2i}=
\frac{\sqrt{m_{\textrm{in}(z_{1})}}}{\eta_{2}}
\left(\frac{\psi^{i}_{\textrm{in}}(z)}{\sqrt{m_{\textrm{in}(z)}}}\right)'|_{z=z_{1}}; \quad i=1,2. 
\label{BDDconditions}
\end{eqnarray}
For a nontrivial solution of \eqref{matrixhom} the determinant of $X$ must be zero
\begin{eqnarray}
|X|=0.
\label{trasEq}
\end{eqnarray}
This is a transcendental equation that has been obtained through the analytical process described and allows us to calculate the energies of the bound states, as seen later in the DHs of Sections  \ref{symdist}, \ref{SMorseDist}, \ref{SExpMass}, and \ref{SubSingMass}.
The energy spectra of these analytical bound states are compared with the reflection coefficient poles, which can be calculated using the multi-step methodology of Section \ref{miltistepm}.\\
When the DH of the Section \ref{symdist} (\ref{SMorseDist}) is generalized through a parameter $\delta$ ($\tau$), as in Subsection \ref{SymGaussMass} (\ref{hypmass}), here,  it is no longer possible to obtain analytical spectra so, we estimate the bound states energy spectra only by the method described in Section  \ref{miltistepm}.

On the other hand, from Eq. \eqref{homsys1} we calculate
\begin{eqnarray}
P=-\frac{\Phi_{12}+\psi^{2}_{\textrm{in}}(z_{0})}{\Phi_{11}\psi^{2}_{\textrm{in}}(z_{0})-\Phi_{12}\psi^{1}_{\textrm{in}}(z_{0})} 
e^{\eta_{0}z_{0}}R, 
\quad
Q=\frac{\Phi_{11}+\psi^{1}_{\textrm{in}}(z_{0})}{\Phi_{11}\psi^{2}_{\textrm{in}}(z_{0})-\Phi_{12}\psi^{1}_{\textrm{in}}(z_{0})} 
e^{\eta_{0}z_{0}}R,
\quad
T=(P\psi^{1}_{\textrm{in}}(z_{1})+ Q\psi^{2}_{\textrm{in}}(z_{1}))e^{\eta_{2}z_{1}}.
\end{eqnarray}

\subsection{Second strategy: Multi-Step method} \label{miltistepm}
The reflection coefficient of the DH, defined by the discrete distributions of potential and mass in Eq. \eqref{potmassdis}, can be calculated using a numerical method.
This method, described in Ref. \cite{Valencia} for the ordering $\alpha=\gamma$, solves the Schrödinger Eq. \eqref{SchrEqj} in each interval $[{z_{\textrm{in}}}_{j-1}, {z_{\textrm{in}}}_{j}]$, taking into account the corresponding boundary conditions for the wave function $\psi(z)$  at each point where there is an abrupt finite change in mass and/or potential.
In this work, we also include the case $\alpha \neq \gamma$, thus generalizing the method the one published in Ref. \cite{Valencia}.

In general, any spatial distribution of the potential $V(z)$ and of mass $m(z)$ can be approximated as a succession of abrupt finite steps. 
The DHs in Section \ref{DobHet} are examples of this approach. 
The values of the parameter pair $\alpha$ and $\gamma$ determine the appropriate boundary conditions for each interface. 
In this work, we have used the boundary conditions \eqref{bcondNew} and \eqref{bcondvonR}, which allow the reflection amplitude $R$ of the charge carriers in the DH with discretized $V(z)$ and $m(z)$ distributions to be obtained recursively; first for one interface, then for two interfaces, and so on. The set of poles of the reflection coefficient $R_{c}=|R|^2$, which can be obtained numerically, represents the energy spectrum of the DH.
  
For the above, we consider a potential function and position-dependent effective mass given by Eq. \eqref{potmassdis}, and we calculate the reflection coefficient ($R$) for an arbitrary integer $n$, which must be large enough in the case of continuous distributions. 
For this purpose, we first consider the following: at a point $z_{j}$ where there is a change of mass \eqref{masaf} and potential, the boundary conditions can be written in general as (see Eqs. \eqref{bcondNew} and \eqref{bcondvonR})
\begin{equation}
\psi_{\textrm{in}j}(z_{\textrm{in}j}^{-})=\mu_{\textrm{in}j}\psi_{\textrm{in}j+1}(z_{\textrm{in}j}^{+}); \quad \psi'_{\textrm{in}j}(z_{\textrm{in}j}^{-})=\rho_{\textrm{in}j}\psi'_{\textrm{in}j+1}(z_{\textrm{in}j}^{+}),
\label{gencondt}
\end{equation}
where $\mu_{\textrm{in}j}=\mu_{\textrm{in}j}(m_{\textrm{in}j},m_{\textrm{in}j+1})$ and $\rho_{\textrm{in}j}=\rho_{\textrm{in}j}(m_{\textrm{in}j},m_{\textrm{in}j+1})$ are functions of the `constant' mass at the left ($m_{\textrm{in}j}$) and right ($m_{\textrm{in}j+1}$) of $z_{\textrm{in}j}$. 
$\psi_{\textrm{in}j}$ and $\psi_{\textrm{in}j+1}$ are the functions at the left and right of $z_{\textrm{in}j}$ respectively. The Schrödinger equation in  $[z_{\textrm{in}j-1},z_{\textrm{in}j}]$ is given by Eq. \eqref{SchrEqj} where 
\begin{eqnarray}
\psi_{\textrm{in}j}(z)=A_{\textrm{in}j}e^{ik_{\textrm{in}j}z}+B_{\textrm{in}j}e^{-ik_{\textrm{in}j}z},
\end{eqnarray} 
$k_{\textrm{in}j}=\sqrt{m_{\textrm{in}j}(E-V_{\textrm{in}j})}$. Applying the general boundary conditions \eqref{gencondt} to the functions $\psi_{\textrm{in}j}$ and $\psi_{\textrm{in}j+1}$ at $z_{\textrm{in}j}$ we get the follow system of equations
 
\begin{align}
A_{\textrm{in}j}e^{ik_{\textrm{in}j}z_{\textrm{in}j}}+B_{\textrm{in}j}e^{-ik_{\textrm{in}j}z_{\textrm{in}j}}
&=\mu_{\textrm{in}j} \left( A_{\textrm{in}j+1}e^{ik_{\textrm{in}j+1}z_{\textrm{in}j}}+B_{\textrm{in}j+1}e^{-ik_{\textrm{in}j+1}z_{\textrm{in}j}}\right),
\label{front1} \\
A_{\textrm{in}j}e^{ik_{\textrm{in}j}z_{\textrm{in}j}}-B_{\textrm{in}j}e^{-ik_{\textrm{in}j}z_{\textrm{in}j}}
&=\frac{k_{\textrm{in}j+1}}{k_{\textrm{in}j}}\rho_{\textrm{in}j} \left(A_{\textrm{in}j+1}e^{ik_{\textrm{in}j+1}z_{\textrm{in}j}}-B_{\textrm{in}j+1}e^{-ik_{\textrm{in}j+1}z_{\textrm{in}j}}\right).
\label{front2}
\end{align}
From Eqs. \eqref{front1} and \eqref{front2} we obtain

\begin{eqnarray}
\frac{B_{\textrm{in}j}}{A_{\textrm{in}j}}
=\frac{r_{j,j+1} + \frac{B_{\textrm{in}j+1}}{A_{\textrm{in}j+1}} e^{-2ik_{\textrm{in}j+1}z_{\textrm{in}j}}}
{1+r_{j,j+1} \frac{B_{\textrm{in}j+1}}{A_{\textrm{in}j+1}} e^{-2ik_{\textrm{in}j+1}z_{\textrm{in}j}}} e^{2ik_{\textrm{in}j}z_{\textrm{in}j}};
\quad
r_{j,j+1}= \frac{k_{\textrm{in}j}\mu_{\textrm{in}j}-k_{\textrm{in}j+1}\rho_{\textrm{in}j}}{k_{\textrm{in}j}\mu_j+k_{\textrm{in}j+1}\rho_{\textrm{in}j}}.
\label{baj}
\end{eqnarray}

\begin{itemize}
\item \textbf{$R$ for a single interfaces ($n=1$)}\\
For a single interface, we have two functions $\psi_{\textrm{in}0}(z)=A_{\textrm{in}0}e^{ik_{\textrm{in}0}z}+B_{\textrm{in}0}e^{-ik_{\textrm{in}0}z}$ and $\psi_{\textrm{in}1}(z)=A_{\textrm{in}1}e^{ik_{\textrm{in}1}z}+B_{\textrm{in}1}e^{-ik_{\textrm{in}1}z}$ whose coefficients satisfy the relation \eqref{baj} with $B_{\textrm{in}1}=0$ and $z_{\textrm{in}0}=0$, so
\begin{eqnarray}
R=\frac{B_{\textrm{in}0}}{A_{\textrm{in}0}}=r_{01}.
\end{eqnarray}

\item \textbf{$R$ for two single interfaces ($n=2$)}\\
In this case, we have three functions $\psi_{\textrm{in}0}(z)=A_{\textrm{in}0}e^{ik_{\textrm{in}0}z}+B_{\textrm{in}0}e^{-ik_{\textrm{in}0}z}$, $\psi_{\textrm{in}1}(z)=A_{\textrm{in}1}e^{ik_{\textrm{in}1}z}+B_{\textrm{in}1}e^{-ik_{\textrm{in}1}z}$, and $\psi_{\textrm{in}2}(z)=A_{\textrm{in}2}e^{ik_{\textrm{in}2}z}$ ($B_{\textrm{in}2}=0$). 
Using Eq. \eqref{baj} we have the relationships
\begin{eqnarray}
\frac{B_{\textrm{in}0}}{A_{\textrm{in}0}}
=\frac{r_{01} + \frac{B_{\textrm{in}1}}{A_{\textrm{in}1}} e^{-2ik_{\textrm{in}1}z_{\textrm{in}0}}}{1+r_{01} \frac{B_{\textrm{in}1}}{A_{\textrm{in}1}} e^{-2ik_{\textrm{in}1}z_{\textrm{in}0}}} e^{2ik_{\textrm{in}0}z_{\textrm{in}0}}, \quad
\frac{B_{\textrm{in}1}}{A_{\textrm{in}1}}
=\frac{r_{12} + \frac{B_{\textrm{in}2}}{A_{\textrm{in}2}} e^{-2ik_{\textrm{in}2}z_{\textrm{in}1}}}{1+r_{12} \frac{B_{\textrm{in}2}}{A_{\textrm{in}2}} e^{-2ik_{\textrm{in}2}z_{\textrm{in}1}}} e^{2ik_{\textrm{in}1}z_{\textrm{in}1}}=r_{12}e^{2ik_{\textrm{in}1}z_{\textrm{in}1}}.
\label{baj2}
\end{eqnarray}
Thus, defining $w_{\textrm{in}1}=z_{\textrm{in}1}-z_{\textrm{in}0}$ and $z_{\textrm{in}0}=0$, the reflection coefficient is
\begin{eqnarray}
R=\frac{B_{\textrm{in}0}}{A_{\textrm{in}0}}
=\frac{r_{01} + r_{12} e^{2ik_{\textrm{in}1}w_{\textrm{in}1}}}{1+r_{01} r_{12} e^{2ik_{\textrm{in}1}w_{\textrm{in}1}}}.
\end{eqnarray}

\item \textbf{$R$ for three single interfaces ($n=3$)}\\
Now, we consider that the space is divided by three interfaces, we will have four wave functions whose coefficients will satisfy the relations
\begin{eqnarray}
\frac{B_{\textrm{in}0}}{A_{\textrm{in}0}}
=\frac{r_{01} + \frac{B_{\textrm{in}1}}{A_{\textrm{in}1}} e^{-2ik_{\textrm{in}1}z_{\textrm{in}0}}}{1+r_{01} \frac{B_{\textrm{in}1}}{A_{\textrm{in}1}} e^{-2ik_{\textrm{in}1}z_{\textrm{in}0}}} e^{2ik_{\textrm{in}0}z_{\textrm{in}0}}, \quad
\frac{B_{\textrm{in}1}}{A_{\textrm{in}1}}
=\frac{r_{12} + \frac{B_{\textrm{in}2}}{A_{\textrm{in}2}} e^{-2ik_{\textrm{in}2}z_{\textrm{in}1}}}{1+r_{12} \frac{B_{\textrm{in}2}}{A_{\textrm{in}2}} e^{-2ik_{\textrm{in}2}z_{\textrm{in}1}}} e^{2ik_{\textrm{in}1}z_{\textrm{in}1}}, \quad
\frac{B_{\textrm{in}2}}{A_{\textrm{in}2}} = r_{23}e^{2ik_{\textrm{in}2}z_{\textrm{in}2}}.
\label{baj3}
\end{eqnarray}
such that the reflection coefficient can be expressed in the following manner
\begin{eqnarray}
R=r_{0123}=\frac{B_{\textrm{in}0}}{A_{\textrm{in}0}}=\frac{r_{01}+r_{123}e^{2ik_{\textrm{in}1}w_{\textrm{in}1}}}{1+r_{01}r_{123}e^{2ik_{\textrm{in}1}w_{\textrm{in}1}}},
\end{eqnarray}
where
\begin{eqnarray}
r_{123}=\frac{r_{12}+r_{23}e^{2ik_{\textrm{in}2}w_{\textrm{in}2}}}{1+r_{12}r_{23}e^{2ik_{\textrm{in}2}w_{\textrm{in}2}}}, 
\label{coefref3}
\end{eqnarray}
$w_{\textrm{in}2}=z_{\textrm{in}2}-z_{\textrm{in}1}$, $w_{\textrm{in}1}=z_{\textrm{in}1}-z_{\textrm{in}0}$, $B_{\textrm{in}3}=0$, and $z_{\textrm{in}0}=0$.

\item \textbf{$R$ for $n$ interfaces}\\
The reflection amplitud of the system described by Eq. \eqref{potmassdis} is
\begin{eqnarray}
R=r_{012\dots n}=\frac{B_{\textrm{in}0}}{A_{\textrm{in}0}}=\frac{r_{01}+r_{12\dots n}e^{2ik_{\textrm{in}1}w_{\textrm{in}1}}}{1+r_{01}r_{12\dots n}e^{2ik_{\textrm{in}1}w_{\textrm{in}1}}},
\label{refcoefpol}
\end{eqnarray}
where
\begin{align}
r_{12\dots n}&=\frac{r_{12}+r_{23\dots n}e^{2ik_{\textrm{in}2}w_{\textrm{in}2}}}{1+r_{12}r_{23\dots n}e^{2ik_{\textrm{in}2}w_{\textrm{in}2}}}, \nonumber \\
r_{23\dots n}&=\frac{r_{23}+r_{34\dots n}e^{2ik_{\textrm{in}3}w_{\textrm{in}3}}}{1+r_{23}r_{34\dots n}e^{2ik_{\textrm{in}3}w_{\textrm{in}3}}}, \nonumber \\
\vdots& \nonumber \\
r_{n-1,n}&=\frac{k_{\textrm{in}n-1}\mu_{\textrm{in}n-1}-k_{\textrm{in}n}\rho_{\textrm{in}n-1}}{k_{\textrm{in}n-1}\mu_{\textrm{in}n-1}+k_{\textrm{in}n}\rho_{\textrm{in}n-1}}.
\label{coefref}
\end{align}
In this equation  we have considered $z_{\textrm{in}0}=0$, $w_{\textrm{in}j}=z_{\textrm{in}j}-z_{\textrm{in}j-1}$,  $\psi_{\textrm{in}n}(z)=A_{\textrm{in}n}e^{ik_{\textrm{in}n}z}$ $(B_{\textrm{in}n}=0)$, and $k_{\textrm{in}j}=\sqrt{m_{\textrm{in}j}(E-V_{\textrm{in}j})}$. 
\end{itemize}

This reflection amplitude $R$ (i.e. reflection coefficient $R_{c}=|R|^2$) depends of the constants $\mu_{\textrm{in}j}$ and $\rho_{\textrm{in}j}$ through $r_{j,j+1}$. For example, if we used the KEO $T_{\textrm{vR}}$ \eqref{kineticop} then (see Eq. \eqref{bcondvonR})
\begin{eqnarray}
\mu_{\textrm{in}j}=\frac{m_{\textrm{in}j+1}^{\alpha+1}m_{\textrm{in}j}^{\alpha+\beta+1}+m_{\textrm{in}j}}{m_{\textrm{in}j}^{\alpha+1}m_{\textrm{in}j+1}^{\alpha+\beta+1}+m_{\textrm{in}j+1}}; \quad
\rho_{\textrm{in}j}=\frac{m_{\textrm{in}j}^{\alpha+1}m_{\textrm{in}j+1}^{\alpha+\beta+1}+m_{\textrm{in}j}}{m_{\textrm{in}j+1}^{\alpha+1}m_{\textrm{in}j}^{\alpha+\beta+1}+m_{\textrm{in}j+1}}.
\label{coeffmurho}
\end{eqnarray}
In the case that we choose the values $\alpha=0$ and $\beta=-1$ ($\nu=\eta=0$), i.e. BenDaniel-Duke operator,
\begin{eqnarray}
\mu_{\textrm{in}j}=1; \quad
\rho_{\textrm{in}j}=\frac{m_{\textrm{in}j}}{m_{\textrm{in}j+1}} \quad \Rightarrow \quad r_{j,j+1}= \frac{\frac{k_{\textrm{in}j}}{m_{\textrm{in}j}}-\frac{k_{\textrm{in}j+1}}{m_{\textrm{in}j+1}}}
{\frac{k_{\textrm{in}j}}{m_{\textrm{in}j}}+\frac{k_{\textrm{in}j+1}}{m_{\textrm{in}j+1}}}.
\label{BenDan}
\end{eqnarray}
On the other hand, for the values $\alpha=-1/2$ and $\beta=0$ ($\nu=-1$, $\eta=-3/2$), i.e. Zhu-Kroemer operator,
\begin{eqnarray}
\mu_{\textrm{in}j}=
\rho_{\textrm{in}j}=\frac{m_{\textrm{in}j}^{1/2}}{m_{\textrm{in}j+1}^{1/2}} \quad \Rightarrow \quad r_{j,j+1}= \frac{k_{\textrm{in}j}-k_{\textrm{in}j+1}}{k_{\textrm{in}j}+k_{\textrm{in}j+1}}.
\label{ZhuKro}
\end{eqnarray}
If we used $\alpha=0$ and $\gamma=-2/3$  ($\nu=-2/3$, $\eta=-4/3$), i.e. $T_{L}$ operator, the coefficients $\mu_{\textrm{in}j}$ and $\rho_{\textrm{in}j}$ are given by Eq. \eqref{bcondNew}
\begin{align}
\mu_{\textrm{in}j}=\frac{2m_{\textrm{in}j}+m_{\textrm{in}j+1}}{m_{\textrm{in}j}+2m_{\textrm{in}j+1}}; \quad 
&\rho_{\textrm{in}j}=\frac{5m_{\textrm{in}j}+m_{\textrm{in}j+1}}{m_{\textrm{in}j}+5m_{\textrm{in}j+1}} \nonumber \\
&\quad \Rightarrow \quad 
r_{j,j+1}=
\frac{\frac{k_{\textrm{in}j}}{(m_{\textrm{in}j}+2m_{\textrm{in}j+1})(5m_{\textrm{in}j}+m_{\textrm{in}j+1})} - \frac{k_{\textrm{in}j+1}}{(m_{\textrm{in}j}+5m_{\textrm{in}j+1})(2m_{\textrm{in}j}+m_{\textrm{in}j+1})}}
{\frac{k_{\textrm{in}j}}{(m_{\textrm{in}j}+2m_{\textrm{in}j+1})(5m_{\textrm{in}j}+m_{\textrm{in}j+1})} + \frac{k_{\textrm{in}j+1}}{(m_{\textrm{in}j}+5m_{\textrm{in}j+1})(2m_{\textrm{in}j}+m_{\textrm{in}j+1})}}.
\label{Lima3}
\end{align}

So, regardless of the value of the coefficients $\mu_{\textrm{in}j}$ and $\rho_{\textrm{in}j}$ in the definition of $r_{j,j+1}$, the recursion form of the reflection coefficient $R$, remains. If $m_{\textrm{in}j}=m_{\textrm{in}j+1}$ then $\mu_{\textrm{in}j}=\rho_{\textrm{in}j}=1$ in all cases. 

We have verified our numerical method by reproducing the analytical spectra published by Christiansen and Lima for the orderings, kinetic operators, and mass profiles they considered, as described in Appendix \ref{AppenChL} and in Ref. \cite{LIMA2023115688}.

\section{Double heterostructure models}\label{DobHet}
In this section, we present various DH models, as defined in the corresponding tables.
For each model, three energy spectra are reported; two associated with $\alpha=\gamma$, and one with $\alpha \neq \gamma$. 
The first column of each table indicates the calculation strategy used. 
The energy spectra obtained in this work for the $\alpha \neq \gamma$ are highlighted, while the results for $\alpha=\gamma$ were taken from \cite{Valencia}.
Appendix \ref{SymDist} complements the analytical results for the DHs defined in Tables \ref{tablesys1} and \ref{TFigT5}. 
Tables \ref{TFigT8} and \ref{FigT9} include mass and potential profiles inspired by Ref. \cite{DESOUZADUTRA200025}.

 \subsection{Symmetric distributions of potential and mass} \label{symdist}
Table \ref{tablesys1} defines the DH and presents its different energy spectra. The first row specifies the distributions according to the notation of equation \eqref{modelmass} and includes their corresponding graphs.
The table presents the energy spectra for three different ambiguities; the first two correspond to the case $\alpha=\gamma$ and the third to a case $\alpha\neq \gamma$.
The spectra with $\alpha=\gamma$ were taken from \cite{Valencia}, and the spectra highlighted in the table corresponding  $\alpha \neq \gamma$ are a result of this work.
The spectra obtained through the solution of the transcendental equation \eqref{trasEq} and those denoted by $\mathbf{E}_{n}$ are explained in the Appendix \ref{SymDist1}.

\subsubsection{Symmetric-Gaussian proﬁles}\label{SymGaussMass}
We now consider some modifications in the DH considered in the Table \ref{tablesys1}. 
Table \ref{TSymmGauss1} (\ref{TSymmGauss2}) defines a family of DHs and reports different energy spectra. 
The first row considers a single potential (mass) for different mass (potential) distributions, characterized through the parameter $\delta$. 
This row also includes some graphs of the masses (potentials), and the effect of parameter $\delta$.\\
Because analytical solutions are difficult to obtain, we analyze these systems numerically. Thus, the energy spectrum is calculated using only the poles of the reflection coefficient \eqref{refcoefpol} for the different ambiguity conditions. 
Tables \ref{TSymmGauss1} and \ref{TSymmGauss2} present the energy spectra where the first two rows correspond to two cases where $\alpha=\gamma$ and the third,  to a case $\alpha \neq \gamma$. The spectra with $\alpha=\gamma$ were taken from \cite{Valencia}, and the spectra corresponding to the case $\alpha \neq \gamma$ are reported in this work.

\subsection{Morse-like distribution of potential and mass}\label{SMorseDist}
Table \ref{TFigT5} defines the DH and presents its different energy spectra. The first row specifies the potential and mass distributions according to
the notation of equation \eqref{modelmass} and includes their corresponding graphs. 
The table presents the energy spectra for three different ambiguities; the first two correspond to the case $\alpha=\gamma$ and the third to a case $\alpha \neq \gamma$.
The spectra with  $\alpha=\gamma$ were
taken from \cite{Valencia}, and the spectrum highlighted in the table corresponding to $\alpha \neq \gamma$ is a result of this work.
The spectrum obtained through
the solution of the transcendental Eq. \eqref{trasEq} is explained in the Appendix \ref{SymDist2}.

\subsubsection{Truncated Morse potential with hyperbolic mass profile}\label{hypmass}
We now consider some modifications in the DH considered in the Table \ref{TFigT5}. Table \ref{TFigT6} defines a family of DHs and reports different energy spectra. 
The first row considers a single potential for different mass distributions, characterized through the parameter $\tau$. 
This row also includes graphs of the mass distributions, highlighting the effect of  parameter $\tau$.

Because analytical solutions are difficult to obtain, we analyze these systems numerically, meaning we calculate the
energy spectrum only using the poles of the reflection coefficient \eqref{refcoefpol} for the different ambiguity conditions. Table
\ref{TFigT6} presents the energy spectra where the first two correspond to the case $\alpha=\gamma$ and the third, to a case $\alpha \neq \gamma$.
The spectra with $\alpha=\gamma$ were taken from \cite{Valencia}, and the spectrum corresponding to the case $\alpha \neq \gamma$ is a result of this
work.

\subsection{Double parabolic quantum well}\label{DoubPQW}
Table \ref{TPotParab} shows different energy spectra of DH defined by
\begin{eqnarray}
V_{p}(z)=V_{0}
\begin{cases}
1\\
f(z)\\
g(z)\\
1
\end{cases}, \quad 
m_{p}(z)=\begin{cases}
m_{0}  &; z<b, \\
m_{1}+(m_{0}-m_{1})f(z) &; b\leq  z < c, \\
m_{1}+(m_{0}-m_{1})g(z) &; c\leq  z < d, \\
m_{0} &; d\leq z.
\end{cases}
\label{parabPot}
\end{eqnarray}
where $f(z)=\frac{(z-c_{1})^2}{c_{2}^2}$,
$g(z)=\frac{(z-c_{3})^2}{c_{4}^2}$;
$c_{1}=\frac{c+a}{2}, \quad c_{2}=\frac{c-a}{2}, \quad c_{3}=\frac{d+c}{2}, \quad c_{4}=\frac{d-c}{2}$.
The first row in  Table \ref{TPotParab} displays their graphs.
Similar to the cases where an analytical solution is not available, the energy spectrum was determined numerically through the poles of the reflection coefficient \eqref{refcoefpol}. The results for the cases $\alpha=\gamma$ were taken from \cite{Valencia}, while the values for $\alpha \neq \gamma$ are newly obtained in the present work.

\subsection{Exponential potential and mass profiles} \label{SExpMass}
\begin{figure}\centering
\includegraphics[scale=0.5]{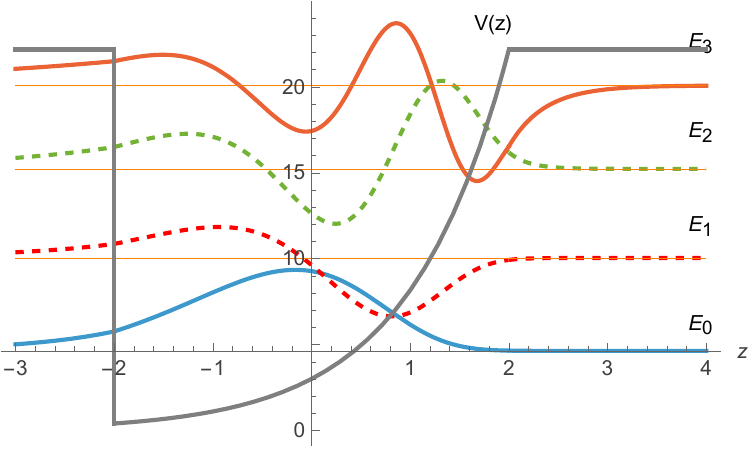}
\caption{Potential $V(z)$ (gray line) of DH \eqref{FigT8} and its eigenfunctions; $\psi^{1}_{\textrm{in}}(z)$ and $\psi^{2}_{\textrm{in}}(z)$ are given by Eq. \eqref{FWitt44}. 
The energy spectrum  $\{E_{0}, E_{1}, E_{2}, E_{3}\}=\{4.63268,10.0389,15.223,20.076\}$ which was obtained using Eq. \eqref{trasEq} corresponds to the Z-K conditions which Table \ref{TFigT8}. Note that the obtained spectrum is discrete and positive, just as the isotonic spectrum \eqref{EnFigT8} is, consistent with Table \ref{PotsIsos}; $V_{c}=3$, $\mu_{0}=1/2$, $c=1$, $\lambda=1$, $z_{0}=-2$, $z_{1}=2$.}
\label{Fig1Whit}
\end{figure}

In this section, we examined a modified profile studied in \cite{DESOUZADUTRA200025}. For our DH model \eqref{modelmass} we define
\begin{eqnarray}
V(z)=
\begin{cases}
V_{0}=\lambda V_{2} \\
V_{\textrm{in}}(z)=V_{c}e^{cz} \\
V_{2}=V_{\textrm{in}}(z_{1})
\end{cases}, \quad
m(z)=
\begin{cases}
m_{0}=m_{\textrm{in}}(z_{0})      &; \quad z\leq z_{0}, \\
m_{\textrm{in}}(z)=\mu_{0}e^{cz} &; \quad z_{0}< z < z_{1}, \label{hetmodel}\\
m_{2}=m_{\textrm{in}}(z_{1})     &; \quad z_{1}\leq z.
\end{cases}
\label{FigT8}
\end{eqnarray}
$V_{c},\mu_{0}>0$. 
It is noteworthy that \( V_{0} = \lambda V_{2} \); $\lambda \in \mathbb{R}$.
The potential distribution  $V (z)$ is illustrated in Fig. \ref{Fig1Whit}.
The corresponding Eq. \eqref{SchomassC} in the inner region $z_{0}<z<z_{1}$ is  
\begin{eqnarray}
-\frac{d^2\phi}{d\rho^2}+ \widetilde{V}(\rho)\phi=E\phi;
\quad \rho=\frac{2\sqrt{\mu_0}}{c}e^{cz/2}>0,
\label{IsotOsc}
\end{eqnarray}
$g = \frac{(3 + 8 \eta - 8 \nu)}{2}$, $\omega^2 = \frac{c^2 V_{c}}{\mu_{0}}$, and
\begin{eqnarray}
\widetilde{V}(\rho)=\frac{1}{4}\omega ^2\rho ^2 +\frac{1}{2}\frac{g}{\rho ^2}.
\label{IsPot}
\end{eqnarray} 
We would like to make the following comments on this potential to clarify some conclusions that were made in Ref. \cite{DESOUZADUTRA200025}.
The parameters $V_{c}$, $c$ and $\mu_{0}$  of the distributions \eqref{FigT8} define the positive parameter $\omega$. 
Independently, the ambiguity parameters $\alpha$ and $\gamma$ determine $g$. 
When $g$ is greater or less than zero, the potential takes the forms described in Table \ref{PotsIsos}.
In principle, both potentials are of a completely different nature:
\begin{itemize}
\item 
The term $g/\rho^2$ is a repulsive (attractive) potential barrier with a positive (negative) singularity  at the origin only when its coefficient $g > 0$ ($g<0$). 
Nevertheless, this term retains the isochronous property that characterizes the harmonic oscillator when $g \geq -1/2$. 
This restriction includes potentials with  $-1/2 \leq g<0$, where the intensity of the centripetal coupling is not strong enough to overcome the harmonic potential barrier.
The energy spectrum of this isotonic potential is
\cite{Calogero,DZhu_1987,Cariñena_2008,Ikhdair,Sesma_2010}.
\begin{eqnarray}
E=\omega  \left(2n+1+d\right); \quad n=0,1,\dots
\label{EnFigT8}
\end{eqnarray}
where
\begin{eqnarray}
d=\frac{1}{2}\sqrt{1+2g}; \quad g\geq -1/2.
\end{eqnarray}
which, as we see, is discrete and positive. 
In fact, $d$ is basically a constant for the factorization of the Hamiltonian in \eqref{IsotOsc} and which according to the references \cite{PhysRevA.57.2851,DODONOV1974597,Nagiyev2006}  is real and positive.
It is straightforward to verify that the operators of G-W, and $T_{L}$ lead to negative $g$ values less than $-1/2$.
Therefore, the graded or double heterostructure that uses these KEOs does not have the energy spectrum \eqref{EnFigT8}. 
For this reason, conclusions about these KEOs used for \eqref{FigT8} cannot be drawn from this energy spectrum.
\item
On the other hand, when $g<-1/2$, the isotonic term ceases to be so.
The potential becomes more attractive by modifying the singularity at the origin in such a way that it must be addressed with special methods for its regularization, with which diverse energy spectra are obtained \cite{Fernández_2021,khelashvili2024}.  
These will be the spectra that must be related to the energy spectra of the heterostructure \eqref{FigT8} when the G-W and $T_{L}$ operators are considered. 
This matter will be detailed in another work. 
\end{itemize}
The Table \ref{PotsIsos}, highlights that the graph of the isotonic potential with $-1/2\leq g <0$  qualitatively coincides with the graph of the non-isotonic potential $g<-1/2$ but their energy spectra are completely different.
The only KEOs with proper name, which define a $g \geq -1/2$, are the Z-K, L-K ($g=-1/2$) and BD-D ($g=3/2$) operators, and therefore the other ambiguities in \eqref{EnFigT8} should not be used.

\begin{table}[h] \centering
\begin{tabular}{lcll}
\toprule
\multicolumn{1}{c}{}                      & $\widetilde{V}(\rho)$                & \multicolumn{1}{c}{Energy spectrum}   & \multicolumn{1}{c}{$g$ value for each KEO}     \\ \hline
\multicolumn{1}{l|}{}                     & \multirow{6}{*}{\includegraphics[scale=0.15]{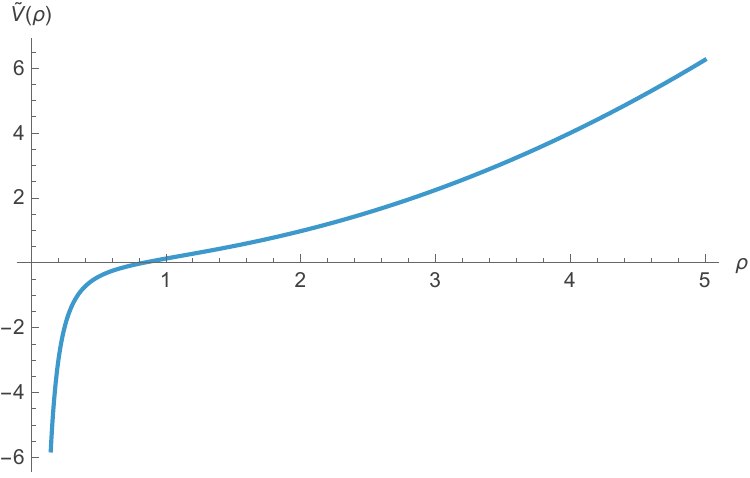}} &                                       &                                                \\
\multicolumn{1}{c|}{$g<-1/2$}             &                         & \multicolumn{1}{c}{Non-isotonic}      & \multicolumn{1}{c}{$- 5/2$ (G-W), $- 7/6$ (L)} \\
\multicolumn{1}{l|}{}                     & \multicolumn{1}{l}{}    &                                       &                                                \\
\multicolumn{1}{l|}{}                     &                                                       &                                       &                                                \\
\multicolumn{1}{c|}{$-1/2 \leq g < 0$} &                         & \multicolumn{1}{c}{Isotonic Eq. \eqref{EnFigT8}} & \multicolumn{1}{c}{$-1/2$ (Z-K and L-K)}       \\
\multicolumn{1}{l|}{}                     & \multicolumn{1}{l}{}    &                                       &                                                \\
\multicolumn{1}{l|}{}                     & \multirow{2}{*}{\includegraphics[scale=0.15]{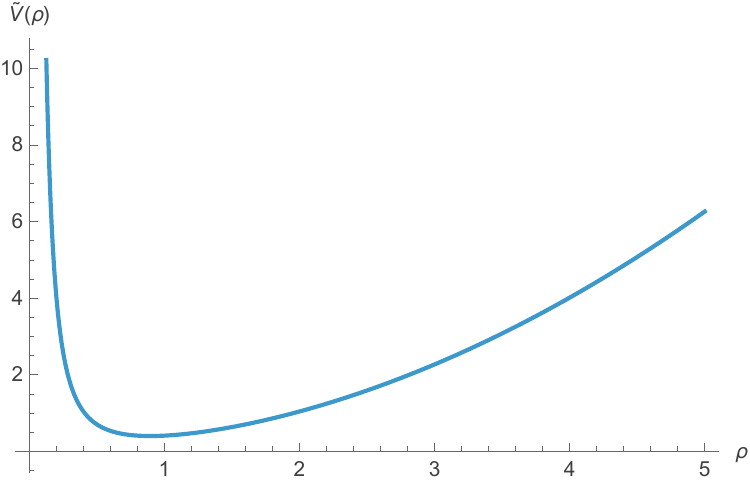}} &                                       &                                                \\
\multicolumn{1}{c|}{$0\leq g$}                &                         & \multicolumn{1}{c}{Isotonic Eq. \eqref{EnFigT8}}  & \multicolumn{1}{c}{$3/2$ (BD-D)}               \\
\multicolumn{1}{l|}{}                     & \multicolumn{1}{l}{}    &                                       &                                                \\ \hline
\end{tabular}
\caption{Potential $\widetilde{V}(\rho)$ \eqref{IsPot} for each of the KEO ambiguities.
The potential with $-1/2 \leq g <0$ is isotonic because it has the energy spectrum \eqref{EnFigT8}, although, qualitatively, its graph corresponds to that of the non-isotonic potential ($g < -1/2$).}
\label{PotsIsos}
\end{table}

We find the solutions of \eqref{IsotOsc} using the ansatz  $\phi=y^{-1/4} F(y)$, where $F(y)$ satisfy the Whittaker equation
\begin{eqnarray}
\frac{d^2 F}{dy^2} +  \left(-\frac{1}{4}+\frac{\kappa}{y}+\frac{1/4-\mu^2}{y^2}\right) F = 0; \quad y=\frac{\omega}{2} \rho^2,
\label{Whitt44}
\end{eqnarray}
$\kappa=E/(2\omega)$, $\mu^2=\frac{1}{16}(1+2g)=\frac{1}{4} (1+2 \eta -2 \nu)$.
Taking into account the Whittaker solutions of first and second kind  $M_{\kappa,\mu}(y)$, $W_{\kappa,\mu}(y)$ of \eqref{Whitt44} respectively, we find \eqref{transf}
\begin{align}
\psi_{\text{in}}^{1}(z) =  M_{\kappa,\mu}(y);
\quad
\psi_{\text{in}}^{2}(z) = W_{\kappa,\mu}(y); \quad y=\frac{2\sqrt{V_{c} \mu_{0}}}{c} e^{cz}. \label{FWitt44}
\end{align}
In Figure \ref{Fig1Whit}, we present the first four non-normalized eigenfunctions and their corresponding energies, derived from Eq. \eqref{trasEq} and Z-K boundary condition. The comprehensive energy spectrum for the KEOs investigated in this study is depicted in Table \ref{TFigT8}.
In the cases where both calculation techniques were used, we observed consistent energy spectra.
An important remark is that the numerical energy spectra depends on the parameter value $V_{0,1}$ as well as the values $z_0$ and $z_1$. These energy values are real for all KEOs of this work.
Choosing some appropriate values of the parameters $z_{0}=-5$, $z_{1}=4$, and $\lambda=1$, the numerical energy values agree with Eq. \eqref{EnFigT8} for the BD-D condition, that is to say, $g=3/2 \Rightarrow E:=E^{BD-D}=w(2n+2)$.

We have verified that when the heterostructure ceases to be double, that is $|z_{0,1}|\rightarrow \infty$ and $V_{0}\rightarrow \infty$, and because of this $0<\rho<\infty$ (Eq. \eqref{IsotOsc}), the energy spectrum of the system \eqref{FigT8} that matches the spectrum of the isotonic oscillator \eqref{EnFigT8} is that of the ambiguity defined by the KEO of BD-D.

\subsection{Singular potential and parabolic mass} \label{SubSingMass}
\begin{figure}\centering
\begin{subfigure}[b]{0.35\linewidth}
\includegraphics[scale=0.5]{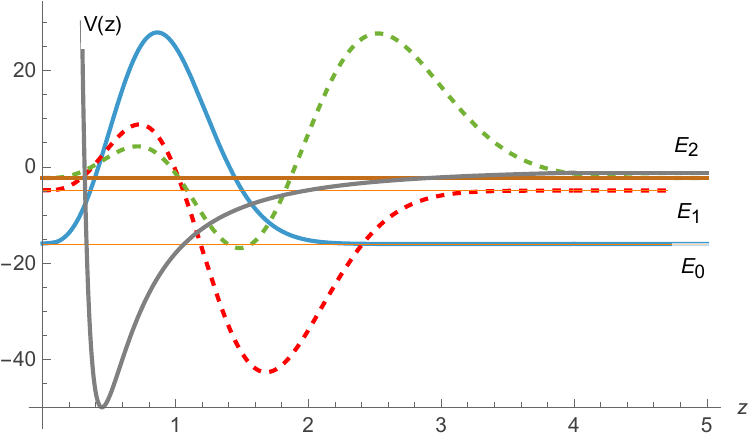}
\caption{Potential $V(z)$ of DH \eqref{singmass}. Conditions  $ A,-B>0$ must be satisfied to generate a quantum well.}
\label{FigFig41}
\end{subfigure}\hspace{10mm}
\begin{subfigure}[b]{0.35\linewidth}
\includegraphics[scale=0.5]{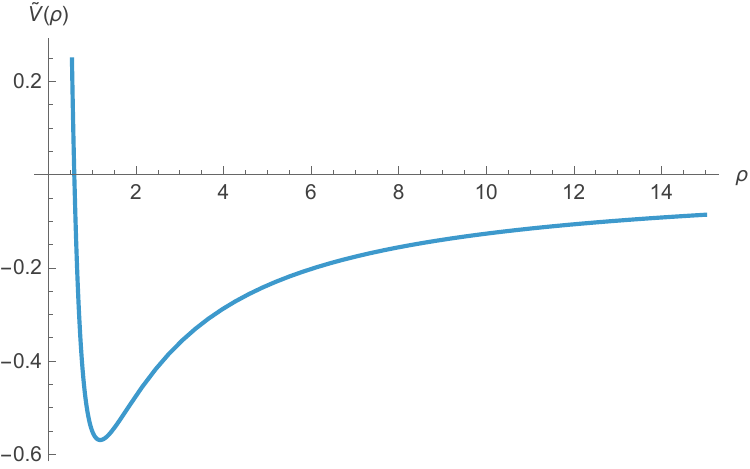}
\caption{Potential $\widetilde{V}(\rho)$ of Hamiltonian \eqref{Potsing2}. Conditions  $F,-G>0$ must be satisfied to generate a quantum well.}
\label{PotSing2}
\end{subfigure}
\caption{a) First three eigenfunctions \eqref{gensol} of the system \eqref{singmass}, $\psi_{\text{in}}^{1}(z)$, $\psi_{\text{in}}^{1}(z)$ are given by Eqs. \eqref{FWitt45} and its energy values $\{E_{0},E_{1},E_{2}\}=\{-16.058849,-4.948710,-2.370368\}$ correspond to Z-K condition; $A=2$, $B=-10$, $c=1$, $z_{0}=0.1$, $z_{1}=4$.}
\end{figure}

Finally, we examined the second profile studied in \cite{DESOUZADUTRA200025} given by
\begin{eqnarray}
m_{\textrm{in}}(z)=c z^2;\quad
V_{\textrm{in}}(z)=\frac{A}{c} \left(\frac{1}{z^4}+\frac{B}{z^2}\right), \quad z,c,A,-B>0.
\label{singmass}
\end{eqnarray}
We defined $V_{0,2}=V_{\textrm{in}}(z_{0,1})$, $m_{0,2}=m_{\textrm{in}}(z_{0,1})$ for our model \eqref{modelmass}.
In the inner region we get  the associated equation \eqref{SchomassC}
\begin{eqnarray}
-\frac{d^2\phi}{d\rho^2}+ \left(\frac{F}{\rho^2}+\frac{G}{\rho}-E\right)\phi=0; \quad \rho=\frac{\sqrt{c}}{2}z^2,
\label{Potsing2}
\end{eqnarray}
$F=(5+8\eta-4\nu+4A)/16$, $G=AB/(2\sqrt{c})$, and $E=-|E|$.
The conditions $F>0$ and $G<0$ must be satisfied to generate a quantum well as the Fig. \ref{PotSing2}.
The first of these conditions links the ambiguity parameters with the potential parameter $A$, that is,
\begin{equation}
A>-\frac{5}{4}-(2\eta-\nu).
\label{Acond}
\end{equation}
This represents a necessary condition that must be considered in advance, before drawing any conclusions based on the energy spectrum of this well.
Eq. \eqref{Potsing2} can be rewritten as the Whittaker equation
\begin{eqnarray}
\frac{d^2\phi}{dy^2} +  \left(-\frac{1}{4}+\frac{\kappa}{y}+\frac{1/4-\mu^2}{y^2}\right) \phi = 0,
\end{eqnarray}
where $y=2\sqrt{|E|}\rho$, $\kappa=-G/(2\sqrt{|E|})$, $\mu^2=\frac{1}{4}+F$. 
Whose solutions are the Whittaker functions of the first and second kind $M_{\kappa,\mu}(y)$, $W_{\kappa,\mu}(y)$ respectively. So, the exact solutions \eqref{transf} in the inner region are
\begin{align}
\psi_{\text{in}}^{1}(z) = c^{1/4}\sqrt{z} M_{\kappa,\mu}(y);
\quad
\psi_{\text{in}}^{2}(z) = c^{1/4}\sqrt{z} W_{\kappa,\mu}(y). \label{FWitt45}
\end{align}
The boundary conditions consistent with the potential well are $\psi(z=0,\infty)=0$.
In Fig. \ref{FigFig41}, we plot the complete eigenfunctions ($-\infty<z<\infty$) of our model \eqref{modelmass} for some parameter values.

We have verified that when the heterostructure ceases to be double and is reduced to the GH of the Ref. \cite{DESOUZADUTRA200025}, that is, in the limit $z_{0}\rightarrow 0$ (i.e. $V_{0}\rightarrow \infty$) and $z_{1}\rightarrow \infty$ $\Rightarrow 0<\rho<\infty$, the analytical energy spectrum of the potential \eqref{Potsing2} is (see Eq. \eqref{enerIsOs} in Appendix \ref{AppendB})
\begin{eqnarray}
E=-\frac{(AB)^2}{4c \left(2n+1+\frac{1}{2}\sqrt{1+2g}\right)^2}; \quad g=2(2+2\eta-\nu)+2A, \quad n=0, 1,\dots
\label{EFigT9}
\end{eqnarray}
Evidently, the sufficient condition to guarantee the reality of the energy spectrum is
\begin{eqnarray}
g>-\frac{1}{2}.
\label{gcond}
\end{eqnarray}
Note that if the ambiguity parameters meet the sufficient condition \eqref{gcond} but not the necessary condition \eqref{Acond}, the energy spectrum \eqref{EFigT9} ceases to make sense.
So, for all conditions, we can always choose an appropriate value for $A>-\frac{5}{4}-(2\eta-\nu)>-\frac{9}{4}-(2\eta-\nu)$ such that the energy spectrum \eqref{EFigT9} is real for all ambiguity
\begin{eqnarray}
{\mathbf{E}_{n}}^{BD-D}=-\frac{(AB)^2}{4c \left(2n+1+\sqrt{9/4+A}\right)^2},
\quad
{\mathbf{E}_{n}}^{Z-K}=-\frac{(AB)^2}{4c \left(2n+1+\sqrt{1/4+A}\right)^2}.
\label{EnSinPot}
\end{eqnarray}
We can observe that the parameter values associated with the $T_{L}$ KEO ($\nu=-2/3, \eta=-4/3$) lead us to the energy spectrum ${\mathbf{E}_{n}}^{T_{L}}={\mathbf{E}_{n}}^{Z-K}$ in Eq. \eqref{EFigT9}.
This observation is only applicable to the GH of the Ref.  \cite{DESOUZADUTRA200025}  and not to the case of DH \eqref{singmass}, where the boundary conditions at points $z_0$ and $z_1$ for each of these KEO are different. 
Therefore, the energy spectra for these KEOs are different, as we can see in Table \ref{FigT9}. 

\section{Conclusions}\label{concl}
In this work, we have generalized the strategies previously proposed in Ref. \cite{Valencia} to calculate the energy spectra in double heterostructures, now including cases where the ambiguity parameters of the kinetic energy operator satisfy $\alpha \neq \gamma$ and the conservation of probability flow at an interface is not guaranteed. 
That is, we have added a case of ambiguity that had not been considered before. 
There are two reasons why we consider that the multi-step method described in Section \ref{miltistepm} is an appropriate procedure for addressing ambiguities in DHs: 1) We have validated both strategies in \cite{Valencia}, showing that the energy spectra of the double heterostructures that admit both procedures are practically identical, and 2) for all the DHs reported here, the numerical multi-step method is sensitive to the different ambiguities considered
and that the orderings $\alpha \neq \gamma$  report finite and well-defined values close to those of the ordering $\alpha = \gamma$.
In most of the models studied (Tables \ref{tablesys1}, \ref{TSymmGauss1}, \ref{TSymmGauss2}, \ref{TFigT5}, \ref{TFigT6}, and \ref{FigT9}), the Zhu-Kroemer (Z-K) operator leads to the lowest spectra, while in other specific cases (Tables \ref{TPotParab} and \ref{TFigT8}), the BD-D and L operators are the most lowest.
We have used the ordering of the kinetic energy operator proposed in reference \cite{lima} and have also shown under what circumstances von Roos's KEO includes it.
Recognizing this connection allows us to unify the treatment of different kinetic energy operators within a common theoretical framework, which is essential for the subsequent analysis of the energy spectra in double heterostructures.
We have not found double heterostructures with ambiguities that prevent obtaining well-defined energy spectra, contrary to what has been reported in other references. For example, Ref. \cite{PhysRevB.39.12783} establishes $\alpha=\gamma=0$ as the universally consistent ordering for KEO in effective mass theory, ensuring physically meaningful spectra for particle jumping in mass within a finite one-dimensional box. 
According to our results, we consider that the conclusions made in this reference are no longer valid for the type of double heterostructures  that we have adopted in this work.
Finally, a critical review has been carried out of the arguments used in the literature to discard certain KEOs, showing that the selection should be based on a detailed analysis of the spectra and boundary conditions, rather than on universal criteria.

\section*{Acknowlegments}
RVT and ECO acknowledges Secretar\'ia de Ciencia, Humanidades, Tecnolog\'ia e Innovaci\'on (SECIHTI - M\'exico) support under the grant FORDECYT-PRONACES/61533/2020. JGR acknowledges IPN project SIP 20260578 and SECIHTI. JA acknowledges IPN project SIP20260576 and SECIHTI.
We extend our sincere gratitude to the referees for their constructive and insightful feedback.

\begin{appendices} 
\setcounter{equation}{0}
\numberwithin{equation}{section}

\section{Boundary conditions for Hamiltonian $H_{L}$}\label{AppenConFronA}
The corresponding Schrödinger equation is

\begin{equation}
H_{L}\psi =
-\frac{\hbar^{2}}{6}
\left[
m^{-1}\frac{d^{2}}{dz^{2}}\psi
+\frac{d}{dz}
\left(
m^{-1}\frac{d}{dz}\psi
\right)
+\frac{d^{2}}{dz^{2}}
\left(
m^{-1}\psi
\right)
\right]
+V(z)\psi=E\psi,
\label{eq:1}
\end{equation}
such that both, the mass and the potential, have a finite discontinuity at
$z_0$, that is\footnote{$H(z-z_0)$ is the Heaviside distribution whose derivative is the Dirac delta distribution.}
\begin{equation}
m^{-1}(z)=
m_{1}^{-1}(z)H(z_{0}-z)
+m_{2}^{-1}(z)H(z-z_{0}),
\label{eq:2}
\end{equation}

\begin{equation}
V(z)=
V_{1}(z)H(z_{0}-z)
+V_{2}(z)H(z-z_{0}),
\label{eq:3}
\end{equation}
where $m_{1}^{-1}(z)$, $m_{2}^{-1}(z)$, $V_{1}(z)$ and
$V_{2}(z)$ are all continuous. 

Therefore, expression~\eqref{eq:1} is an equation in the sense of
distributions whose solutions could have a finite discontinuity at
$z_0$,

\begin{equation}
\psi(z)=
\phi_{1}(z)H(z_{0}-z)
+\phi_{2}(z)H(z-z_{0}).
\label{eq:4}
\end{equation}
$\phi_{1,2}(z)\in W_2^2(\mathbb{R})$, where this is the set of continuous
functions that possess a continuous derivative at every point and whose
second derivative exists almost everywhere. In this way, the domain of
the Hamiltonian, $H_{L}$, will be $W_2^2(\mathbb{R}/z_0)$. 
The functions in this
space satisfy the same properties as the functions in $W_2^2(\mathbb{R})$
except that they and their derivatives can admit a finite jump in $z_0$.
Differentiating~\eqref{eq:4},

\begin{equation}
\begin{aligned}
\psi'(z)
=
\phi_{1}'(z)H(z_{0}-z)
+\phi_{2}'(z)H(z-z_{0})+
\left[
\phi_{2}(z_{0})-\phi_{1}(z_{0})
\right]\delta(z-z_{0}),
\end{aligned}
\label{eq:5}
\end{equation}
differentiating once more,
\begin{equation}
\begin{aligned}
\psi''(z)
={}
\phi_{1}''(z)H(z_{0}-z)
+\phi_{2}''(z)H(z-z_{0})
+\left[
\phi_{2}'(z_{0})-\phi_{1}'(z_{0})
\right]\delta(z-z_{0})+
\left[
\phi_{2}(z_{0})-\phi_{1}(z_{0})
\right]\delta'(z-z_{0}).
\end{aligned}
\label{eq:6}
\end{equation}
Now we multiply \eqref{eq:6} by $m^{-1}(z)$ 
\begin{equation}
\begin{aligned}
m^{-1}(z)\psi''(z)
={}&
m_{1}^{-1}(z)\phi_{1}''(z)H(z_{0}-z)
+m_{2}^{-1}(z)\phi_{2}''(z)H(z-z_{0})+\\
&+
\left[
\phi_{2}'(z_{0})-\phi_{1}'(z_{0})
\right]
m^{-1}(z)\delta(z-z_{0})
+
\left[
\phi_{2}(z_{0})-\phi_{1}(z_{0})
\right]
m^{-1}(z)\delta'(z-z_{0}),
\end{aligned}
\label{eq:7}
\end{equation}
in this expression, we observe two conflicting products,
$m^{-1}(z)\delta(z-z_{0})$ and
$m^{-1}(z)\delta'(z-z_{0})$. Let us look at the first one,

\begin{equation}
m^{-1}(z)\delta(z-z_{0})
=
\left[
m_{1}^{-1}(z)
+
\left(
m_{2}^{-1}(z)-m_{1}^{-1}(z)
\right)
H(z-z_{0})
\right]\delta(z-z_{0}),
\label{eq:8}
\end{equation}
it involves the product of the distributions $H(z-z_0)$ and
$\delta(z-z_0)$, which is defined within the framework of Schwartz
distribution theory~\cite{DisTheoAppKinani}

\begin{equation}
H(z-z_{0})\delta(z-z_{0})
=
\frac{1}{2}\delta(z-z_{0}),
\label{eq:9}
\end{equation}
hence,
\begin{equation}
m^{-1}(z)\delta(z-z_{0})
=
\frac{
m_{1}^{-1}(z_{0})+
m_{2}^{-1}(z_{0})
}{2}
\delta(z-z_{0}).
\label{eq:10}
\end{equation}
The second product, $m^{-1}(z)\delta'(z-z_0)$, is more delicate
\begin{equation}
m^{-1}(z)\delta'(z-z_{0})
=
\left[
m_{1}^{-1}(z)
+
\left(
m_{2}^{-1}(z)-m_{1}^{-1}(z)
\right)
H(z-z_{0})
\right]\delta'(z-z_{0}),
\label{eq:11}
\end{equation}
this involves the product of distributions $H(z-z_0)$ and
$\delta'(z-z_0)$, as well as the product of continuous functions with
the distribution $\delta'(z-z_0)$. The latter case is well-defined in
the Schwartz distribution theory~\cite{DisTheoAppKinani}
\begin{equation}
f(z)\delta'(z-z_{0})
=
f(z_{0})\delta'(z-z_{0})
-f'(z_{0})\delta(z-z_{0}).
\label{eq:12}
\end{equation}
However, the former is problematic in the sense that the square of the
Dirac delta function arises~\cite{CKRaju1982,KohKuanChen}
\begin{equation}
H(z-z_{0})\delta'(z-z_{0})
=
\frac{1}{2}\delta'(z-z_{0})
-\delta^{2}(z-z_{0}).
\label{eq:13}
\end{equation}

There is still controversy surrounding the definition of powers of the
Dirac delta; it is necessary to resort, among other concepts, to
fractional derivatives~\cite{LI2014502}, conformable derivatives~\cite{Jarad},
and neutrix limits~\cite{LI2014502,Jarad,Emin} in order to define them.
Within these schemes, any power of the Dirac delta distribution is zero, $\delta^k (z-z_0 )=0$ if $k$ is even. 
However, in the Colombeau algebra, the most
rigorous framework for handling the product of distributions, the square
of the Dirac delta generally remains infinite, vanishing only in very
specific cases~\cite{Colombeau1990,colombeau1984new,AguideCham}. Consequently, the term involving
the square of the Dirac delta remains highly problematic and generally
leads to controversial physical results.

Using~\eqref{eq:9}, \eqref{eq:12}, and \eqref{eq:13}, we find
\begin{equation}
\begin{aligned}
m^{-1}(z)\delta'(z-z_{0})
={}
\frac{1}{2}
\left[
\frac{m_{1}'(z_{0})}{m_{1}^{2}(z_{0})}
+
\frac{m_{2}'(z_{0})}{m_{2}^{2}(z_{0})}
\right]\delta(z-z_{0})+\\
+
\frac{
m_{1}^{-1}(z_{0})+
m_{2}^{-1}(z_{0})
}{2}\delta'(z-z_{0})
&+
\left[
m_{1}^{-1}(z)-m_{2}^{-1}(z)
\right]\delta^{2}(z-z_{0}).
\end{aligned}
\label{eq:14}
\end{equation}
Substituting~\eqref{eq:10} and~\eqref{eq:14} into~\eqref{eq:7},
\begin{equation}
\begin{aligned}
m^{-1}(z)\psi''(z)
&={}
\frac{1}{m_{1}(z)}
\phi_{1}''(z)H(z_{0}-z)
+
\frac{1}{m_{2}(z)}
\phi_{2}''(z)H(z-z_{0})+\\
&+
\left[
\left(
\phi_{2}'(z_{0})-\phi_{1}'(z_{0})
\right)
\frac{
m_{1}^{-1}(z_{0})+
m_{2}^{-1}(z_{0})
}{2}
+
\left(
\phi_{2}(z_{0})-\phi_{1}(z_{0})
\right)
\left(
\frac{m_{1}'(z_{0})}{2m_{1}^{2}(z_{0})}
+
\frac{m_{2}'(z_{0})}{2m_{2}^{2}(z_{0})}
\right)
\right]\delta(z-z_{0})+\\
&+
\left(
\phi_{2}(z_{0})-\phi_{1}(z_{0})
\right)
\frac{
m_{1}^{-1}(z_{0})+
m_{2}^{-1}(z_{0})
}{2}
\delta'(z-z_{0})+\\
&+
\left(
\phi_{2}(z_{0})-\phi_{1}(z_{0})
\right)
\left[
m_{1}^{-1}(z)-m_{2}^{-1}(z)
\right]
\delta^{2}(z-z_{0}).
\end{aligned}
\label{eq:15}
\end{equation}

In what follows we will keep the terms with $\delta^2(z-z_0)$ and see
that this leads to some inconsistencies. The remaining two terms of the
kinetic energy operator in equation~\eqref{eq:1} are calculated in a
similar manner.
Thus, the Schrödinger equation~\eqref{eq:1} can be written as

\begin{equation}
\begin{aligned}
&\left[
\frac{d}{dz}\frac{1}{m_{1}(z)}
\frac{d}{dz}\phi_{1}(z)
+
\left(
\frac{2}{3}\frac{[m_{1}'(z)]^{2}}{m_{1}^{3}(z)}
-\frac{1}{3}\frac{m_{1}''(z)}{m_{1}^{2}(z)}
\right)\phi_{1}(z)
-\frac{2}{\hbar^{2}}
\left(V_{1}(z)-E\right)\phi_{1}(z)
\right]H(z_{0}-z)+\\
&+
\left[
\frac{d}{dz}\frac{1}{m_{2}(z)}
\frac{d}{dz}\phi_{2}(z)
+
\left(
\frac{2}{3}\frac{[m_{2}'(z)]^{2}}{m_{2}^{3}(z)}
-\frac{1}{3}\frac{m_{2}''(z)}{m_{2}^{2}(z)}
\right)\phi_{2}(z)
-\frac{2}{\hbar^{2}}
\left(V_{2}(z)-E\right)\phi_{2}(z)
\right]H(z-z_{0})-\\
&-\frac{1}{\hbar^{2}m_{1}(z_{0})m_{2}(z_{0})}
\left[
\phi_{2}'(z_{0})
\left(5m_{1}(z_{0})+m_{2}(z_{0})\right)
-\phi_{1}'(z_{0})
\left(m_{1}(z_{0})+5m_{2}(z_{0})\right)\right.+\\
&\left.
\qquad
+
\left(\phi_{1}(z_{0})+\phi_{2}(z_{0})\right)
\left(
\frac{m_{2}(z_{0})m_{1}'(z_{0})}{m_{1}(z_{0})}
-
\frac{m_{1}(z_{0})m_{2}'(z_{0})}{m_{2}(z_{0})}
\right)
\right]\delta(z-z_{0})-\\
&-\frac{2}{\hbar^{2}m_{1}(z_{0})m_{2}(z_{0})}
\left[
\phi_{2}(z_{0})
\left(2m_{1}(z_{0})+m_{2}(z_{0})\right)
-
\phi_{1}(z_{0})
\left(m_{1}(z_{0})+2m_{2}(z_{0})\right)
\right]\delta'(z-z_{0})-\\
&-\frac{2}{\hbar^{2}}
\left(\phi_{2}(z_{0})-\phi_{1}(z_{0})\right)
\left[
m_{1}^{-1}(z)-m_{2}^{-1}(z)
\right]\delta^{2}(z-z_{0})
=0.
\end{aligned}
\label{eq:16}
\end{equation}

This leads us to the system of equations

\begin{equation}
-\frac{\hbar^{2}}{2}
\frac{d}{dz}
\frac{1}{m_{1}(z)}
\frac{d}{dz}\phi_{1}(z)
+
\left[
\frac{2}{3}
\frac{[m_{1}'(z)]^{2}}{m_{1}^{3}(z)}
-
\frac{1}{3}
\frac{m_{1}''(z)}{m_{1}^{2}(z)}
\right]\phi_{1}(z)
+
V_{1}(z)\phi_{1}(z)
=
E\phi_{1}(z).
\label{eq:17}
\end{equation}

\begin{equation}
-\frac{\hbar^{2}}{2}
\frac{d}{dz}
\frac{1}{m_{2}(z)}
\frac{d}{dz}\phi_{2}(z)
+
\left[
\frac{2}{3}
\frac{[m_{2}'(z)]^{2}}{m_{2}^{3}(z)}
-
\frac{1}{3}
\frac{m_{2}''(z)}{m_{2}^{2}(z)}
\right]\phi_{2}(z)
+
V_{2}(z)\phi_{2}(z)
=
E\phi_{2}(z).
\label{eq:18}
\end{equation}

\begin{equation}
\begin{aligned}
\phi_{2}'(z_{0})
\left(5m_{1}(z_{0})+m_{2}(z_{0})\right)
&-
\phi_{1}'(z_{0})
\left(m_{1}(z_{0})+5m_{2}(z_{0})\right)\\
&+
\left(\phi_{1}(z_{0})+\phi_{2}(z_{0})\right)
\left[
\frac{m_{2}(z_{0})m_{1}'(z_{0})}{m_{1}(z_{0})}
-
\frac{m_{1}(z_{0})m_{2}'(z_{0})}{m_{2}(z_{0})}
\right]
=0.
\end{aligned}
\label{eq:19}
\end{equation}

\begin{equation}
\phi_{2}(z_{0})
\left(2m_{1}(z_{0})+m_{2}(z_{0})\right)
-
\phi_{1}(z_{0})
\left(m_{1}(z_{0})+2m_{2}(z_{0})\right)
=0.
\label{eq:20}
\end{equation}

\begin{equation}
\left[
\phi_{2}(z_{0})-\phi_{1}(z_{0})
\right]
\left[
m_{1}^{-1}(z)-m_{2}^{-1}(z)
\right]
\delta^{2}(z-z_{0})
=0.
\label{eq:21}
\end{equation}
Expression~\eqref{eq:20}, obtained by requiring that the coefficient of
$\delta'(z-z_0)$ in~\eqref{eq:16} be zero, is precisely the boundary
condition for the wave function found in~\cite{lima} and used in our
work. This result is inconsistent with the one we would obtain if we
required that the coefficient accompanying $\delta^2(z-z_0)$ be zero,
where we would find that the wave function would be continuous at the
mass jump, as shown in~\eqref{eq:21}. In order to avoid this
inconsistency, it is necessary to consider that
$\delta^2(z-z_0)=0$, just as mentioned before.\\
Equation~\eqref{eq:19}, obtained by requiring the coefficient of
$\delta(z-z_0)$ in~\eqref{eq:16} to be zero, gives the boundary
condition at the mass jump for the derivative of the wave function.
This result differs from that obtained in~\cite{lima} by a term that
depends on the right-hand and left-hand limits of the mass derivative
on both sides of the mass jump. For the case of constant masses on both
sides of the boundary, this additional term vanishes, and the boundary
condition \eqref{bcondNew} for the derivative obtained in~\cite{lima} is recovered.
This is precisely the situation in our methodology employed in this
work.\\
For this case, constant masses on both sides of the boundary, a simple
calculation of the probability flux on both sides of the boundary leads
us to conclude that such flux is not conserved:
\begin{equation}
J(z_{0}^{+})
=
\frac{
7m_{1}+10m_{2}+\dfrac{m_{1}^{2}}{m_{2}}
}{
7m_{2}+10m_{1}+\dfrac{m_{2}^{2}}{m_{1}}
}
J(z_{0}^{-}).
\label{eq:22}
\end{equation}
For example, when $m_{1}=2m_{2}$, it turns out that
\begin{equation}
J(z_{0}^{+})=1.08J(z_{0}^{-}).
\end{equation}
Setting $\delta^2(z-z_0)=0$ resolved an inconsistency in the boundary
condition for the wave function but does not resolve controversial
physical results, such as the non-conservation of flux. The
non-conservation of flux would imply that the interface where a mass
jump occurs creates or annihilates particles.
This is precisely one of the consequences of the Hamiltonian~\eqref{eq:1}
being ill-conditioned. Nevertheless, this Hamiltonian can still be
useful, specifically when studying interfaces between semiconductors
where particles may be absorbed or emitted.
Strictly speaking, the only way for the Hamiltonian to be free of inconsistencies and become self-adjoint is for $\phi_{2}(z_{0}) = \phi_{1}(z_{0})=0$. 
This situation represents the trivial, uninteresting case in which regions 1 and 2 are isolated; that is, 
the quasiparticles can not cross the boundary and remain within region 1 or 2.

\section{Verification of analytical energy spectra given in Ref. \cite{LIMA2023115688}}\label{AppenChL}

\subsection{Example 3.1: Soliton-like and reciprocal biquadratic profiles}
The soliton-like mass, $m_{s}(z)=\frac{1}{\cosh^{2}{z}}$, and the reciprocal biquadratic mass, $m_{b}(z)=\frac{1}{(1+z^2)^2}$, given by Eqs. (7) in \cite{LIMA2023115688},
give us the effective potential \eqref{Vtilde}
\begin{eqnarray}
\widetilde{V}(\rho)=\omega_{\alpha \gamma}\tan^2{\rho}+\widetilde{V}(0); 
\end{eqnarray}
where
\begin{eqnarray}
\omega_{\alpha \gamma}=
\begin{cases}
4\alpha \gamma+2(\alpha+\gamma)+\frac{3}{4}\\
16\alpha \gamma+6(\alpha+\gamma)+2
\end{cases}
, \quad 
\widetilde{V}(0)=
\begin{cases}
\alpha+\gamma+\frac{1}{2}; \quad \text{for the soliton-like mass},\\
2(\alpha+\gamma)+1;\quad \text{for the reciprocal biquadratic mass}.
\end{cases}
\end{eqnarray}
and
\begin{eqnarray}
\rho(z) =
\begin{cases}
\sinh{z}=\tan{\rho}; \quad \text{for the soliton-like mass},\\
z=\tan{\rho};\quad \text{for the reciprocal biquadratic mass}.
\end{cases}
\end{eqnarray}
According to  \cite{LIMA2023115688}, when $z \in (-\infty,\infty)$ $\Leftrightarrow$ $\rho \in  \left(- \frac{\pi}{2},\frac{\pi}{2}\right)$, the system has bound states only when $-1/4<\omega_{\alpha \gamma} $ and its energy spectrum is
\begin{eqnarray}
E_{n}=n^2+2(1-2\nu_{\alpha \gamma})n+1-2\nu_{\alpha \gamma}+\widetilde{V}(0); 
\quad 
\nu_{\alpha \gamma}=\frac{1}{4} \left(1-\sqrt{1+4\omega_{\alpha \gamma}}\right),
\quad
n \in \mathbb{Z^{+}}.
\label{EenerSB}
\end{eqnarray}
To compare this analytical spectrum with our numerical results, we consider $z \in [z_{0}, z_{1}]$ (with $z_{1} = -z_{0} > 0$), which implies $\rho \in [\rho_{0}, \rho_{1}] \subset \left(-\frac{\pi}{2}, \frac{\pi}{2}\right)$ and the following profiles for our double heterostructure model
\begin{eqnarray}
V(z)=
\begin{cases}
V_{0} \\
0\\
V_{0}
\end{cases}, \quad
m(z)=
\begin{cases}
m_{\textrm{in}}(z_{0})&; \quad z\leq z_{0}, \\
m_{\textrm{in}}(z) &; \quad z_{0}< z < z_{1}, \label{hetmodel5}\\
m_{\textrm{in}}(z_{0})&; \quad z_{1}\leq z.
\end{cases}
\end{eqnarray}
with
\begin{eqnarray}
V_{0}=
\begin{cases}
\frac{3}{4}\tan^2{\rho_{0}}+\frac{1}{2}\\
2\tan^2{\rho_{0}}+1
\end{cases}
, \quad
m_{\textrm{in}}(z)=
\begin{cases}
m_{s}(z); \quad \text{for the soliton-like mass},\\
m_{b}(z); \quad \text{for the reciprocal biquadratic mass}.
\end{cases}
\end{eqnarray}
The election of $V_0$ corresponds to the values for the BDD ambiguity parameters (see Table 1 in \cite{LIMA2023115688}).
The selection of this value for $V_0$ is arbitrary and directly and significantly influences the energy spectrum of the double heterostructure, as well as the values of $z_{0,1}$. Once it has been fixed, we can numerically calculate the energy spectrum corresponding to each KEO studied in our work.
Table \ref{TSoliBiq} presents the reflection coefficients for each of the operators investigated in this article. Furthermore, it includes the numerical results obtained using the MM ($\alpha=\gamma=-1/4$) studied in reference \cite{LIMA2023115688}. 
It is noteworthy that the values associated with the BBD operator align with the analytical results—as anticipated, since the constant $V_0$ is deliberately selected to coincide with the potential $\widetilde{V}(\rho_0)$. In contrast, the ZK operator is disregarded in \cite{LIMA2023115688} when $z \in (-\infty,\infty)$ ($\Leftrightarrow$ $\rho \in  \left(- \frac{\pi}{2},\frac{\pi}{2}\right)$). However, in the finite case, it produces a bound-state spectrum for the soliton mass profile.

\begin{table}[h]
\begin{tabular}{lcccc|clcccc}
\hline
\multicolumn{1}{c}{Soliton-like mass}             & $E_0$   & $E_1$   & $E_2$   & ... &  & \multicolumn{1}{c}{Biquadratic mass}         & $E_0$   & $E_1$   & $E_2$   & ... \\ \hline
$E_n$ (Eq. \eqref{EenerSB}) & 2       & 6       & 12      & ... &  & $E_n$ (Eq. \eqref{EenerSB}) & 3       & 8       & 15      & ... \\
\rowcolor[HTML]{B8B7B7}
BDD                                          & 1.99994 & 5.99967 & 11.999  & ... &  & BDD                                          & 2.9943  & 7.96518 & 14.8809 & ... \\
\rowcolor[HTML]{B8B7B7}
ZK                                           & 2.00232 & 6.00376 & 12.005  & ... &  & ZK                                           & 3.31378 & 8.70339 & 16.2438 & ... \\
\rowcolor[HTML]{B8B7B7}
L                                            & 1.86212 & 5.36527 & 10.7993 & ... &  & L                                            & 3.3913  & 8.4261  & 15.6136 & ... \\
\rowcolor[HTML]{B8B7B7}
MM                                           & 0.868   & 3.46919 & 7.79478 & ... &  & MM                                           & 1.03806 & 4.15126 & 9.336   & ... \\ \hline
\end{tabular}
\caption{Numerical energy spectra of the system \eqref{hetmodel5} obtained by the multi-step method described in Section \ref{miltistepm}.
The BDD values (first row) agree with the analytical spectrum \eqref{EenerSB}.
The spectra are sensitive to both the value $V_0$ and the values of $z_{0,1}$. 
In particular, an appropriate adjustment of $V_0$ would allow matching the values for the MM operator with the corresponding analytical spectrum obtained from \eqref{EenerSB}. 
The highlighted lines are contributions of this article;
$z_{0}=-10=-z_{1}$.}
\label{TSoliBiq}
\end{table}

\subsection{Example 3.2: Reciprocal quadratic mass}
Reciprocal quadratic mass given by Eq. (22) in \cite{LIMA2023115688}, $m_{q}(z)=(1+z^2)^{-1}$,
this is a particular case  of the system given by Eq. \eqref{sym1} ($\sigma=1$, $\mu=0$). 
In Section 3.2.4 the authors say: ``For this mass profile, the ZK ordering is the only one having bound states; more specifically, just the ground state'' because
\begin{eqnarray}
\omega_{\alpha \gamma}=\frac{1}{4}-4\alpha \gamma,
\end{eqnarray}
must satisfy $\omega_{\alpha \gamma}<0$, (see Eq. (24b) and Table 2 in \cite{LIMA2023115688}). However, taking into account the potential $V(z)=-\mu^2/(1+z^2)$ we get
\begin{eqnarray}
\omega_{\alpha \gamma}=\frac{1}{4}-4\alpha \gamma-\mu^2 \sigma^2,
\end{eqnarray}
and we can always choose appropriate values for $\mu$ and $\sigma$ that for all KEO $\omega_{\alpha \gamma}<0$ is satisfied. According to Ref. \cite{LIMA2023115688}, in the particular case, $\sigma=1$, $\mu=0$, when 
$z \in (-\infty,\infty)$ $\Leftrightarrow$ $\rho \in (-\infty,\infty)$, the energy spectrum is
\begin{eqnarray}
E_{n}=-n^2+(2\sqrt{\alpha \gamma}-1)n+2\sqrt{\alpha \gamma}+\alpha+\gamma.
\label{ELimaSpectre}
\end{eqnarray}
This result is consistent with Eq. \eqref{EnerPTpot}.

As in the previous example, for the finite case when $z \in [z_{0}, z_{1}]$ $\Leftrightarrow$ $\rho \in [\rho_{0}, \rho_{1}]$, let us consider the next profile
\begin{eqnarray}
V(z)=
\begin{cases}
V_{0}=-\frac{3}{4} \sech{\rho_{0}}+\frac{1}{4} \\
0\\
V_{0}=-\frac{3}{4} \sech{\rho_{0}}+\frac{1}{4}
\end{cases}, \quad
m(z)=
\begin{cases}
m_{\textrm{in}}(z_{0})&; \quad z\leq z_{0}, \\
m_{\textrm{in}}(z)=m_{q}(z) &; \quad z_{0}< z < z_{1}, \label{hetmodel}\\
m_{\textrm{in}}(z_{0})&; \quad z_{1}\leq z.
\end{cases}
\end{eqnarray}
This double heterostructure model yields the following spectra for each KEO: 
$\{0.1895,0.2275 \}$ for BDD, 
$\{0.0234\}$ for ZK, 
$\{0.0618\}$ for L,
$\{0.0917\}$ for MM.

\section{}\label{SymDist}
In this appendix, the analytical method of Section \ref{AnalyMetho} is applied to the double heterostructures of Sections \ref{symdist} and \ref{SMorseDist}. The energy spectra presented in Tables \ref{tablesys1} and \ref{TFigT5} are explained here.
\subsection{Complement of the DH defined in Table \ref{tablesys1}.} \label{SymDist1}
\begin{figure}[h!]
\centering
\begin{subfigure}[b]{0.35\linewidth}
\includegraphics[width=\linewidth]{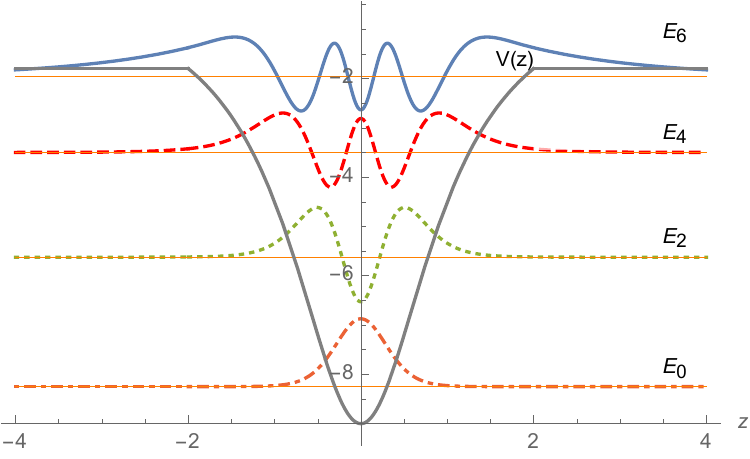}
\caption{Even wave functions}
\label{EvenWF}
\end{subfigure} \hspace{10mm}
\begin{subfigure}[b]{0.35\linewidth}
\includegraphics[width=\linewidth]{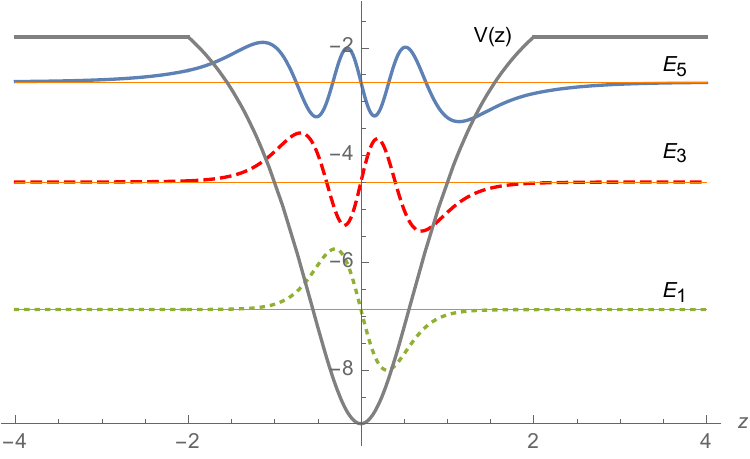}
\caption{Odd wave functions}
\label{OddWF}
\end{subfigure}
\caption{Wave function \eqref{gensol} with $\psi_{\text{in}}^{1}(z)$, $\psi_{\text{in}}^{1}(z)$ given by Eqs. \eqref{solsym1} for the distributions \eqref{sym1}. The energy values, $\{E_{0},E_{1},E_{2},E_{3},E_{4},E_{5},E_{6}\}=\{-8.25, -6.875, -5.625, -4.50009, -3.5013, -2.63724, -1.96428\}$, are found through the trascendental equation \eqref{trasEq} using the BD-D condition.}
\label{FigSymm}
\end{figure}

We study the symmetric DH model with potential and mass distributions in the inner region $z_{0}<z<z_{1}$ in Eq. \eqref{modelmass} 
\begin{eqnarray}
V_{\textrm{in}}(z)=-\frac{\mu^2}{1+z^2}, \quad m_{\textrm{in}}(z)=\frac{\sigma^2}{1+z^2},
\label{sym1}
\end{eqnarray}
$\mu$, $\sigma$ are non-zero real parameters. We define $z_{0}<0$, $z_{1}=|z_{0}|$, $V_{0}=V_{2}=V_{\textrm{in}}(z_{0})$, and $m_{0}=m_{2}=m_{\textrm{in}}(z_{0})$, so that $V(z)$ and $m(z)$ are continuous as is shown in figure inserted in Table \ref{tablesys1}. The associated Schrödinger equation \eqref{SchomassC} for this system can be written as
\begin{eqnarray}
\frac{d^2\phi}{d\rho^2}+ \left(\kappa^2 +\frac{\lambda(\lambda-1)}{\sigma^2} \frac{1}{\cosh^2{\frac{\rho}{\sigma}}}\right)\phi=0;
\quad
\rho=\sigma \sinh^{-1}{z},
\label{kaplamEq}
\end{eqnarray}
where
\begin{eqnarray}
\kappa^2=E-\frac{1/4+2\eta-3\nu}{\sigma^2}, 
\quad
 \lambda(\lambda-1)=-(1/4+4\nu-2\eta-\mu^2\sigma^2).
\label{kaplam}
\end{eqnarray}
For the bound states to exist $\lambda(\lambda-1)>0$. From Eq. \eqref{kaplam} we find $\lambda=\frac{1}{2}+\sqrt{\mu^2\sigma^2+2\eta-4\nu}$. Taking into account the solution $\phi$ of Eq. \eqref{kaplamEq}, we find \cite{Valencia}  the general solution  \eqref{gensol} in the inner region
\begin{align}
\psi^{1}_{\textrm{in}}(z) = A(z)\leftindex_{2}{F}_{1} \left(a,b,\frac{1}{2};-z^2\right),
\quad
\psi^{2}_{\textrm{in}}(z)=  A(z) z  \leftindex_{2}{F}_{1} \left(a+\frac{1}{2},b+\frac{1}{2},\frac{3}{2};-z^2\right);
\label{solsym1}
\end{align}
where $A(z)= \left(\frac{\sigma^2}{1+z^2}\right)^{1/4}(1+z^2)^{\lambda/2}$, $a=\frac{1}{2}(\lambda+i\kappa\sigma)$, $b=\frac{1}{2}(\lambda-i\kappa\sigma)$. Applying these solutions \eqref{solsym1} to the transcendental equation \eqref{trasEq}, we obtain the `analytical' energy spectrum. Fig. \eqref{FigSymm} displays graphs of both odd and even solutions, while Table \eqref{tablesys1} presents the energy spectra obtained for various KEOs.

On the other hand, we can see that Eq. \eqref{kaplamEq} is the Schrödinger equation for a constant mass particle subject to a modified Pöschl-Teller potential. The energy spectrum, $\kappa^2$ (i.e. $E:=\mathbf{E}_{n}$), and its solutions \cite{flugge,Valencia} in the whole space is well known, that is to say, if $z_{0}\rightarrow -\infty$, $z_{1}\rightarrow \infty$ in the model \eqref{modelmass} we find
\begin{eqnarray}
\mathbf{E}_{n}= -\frac{(n-\lambda+1)^2}{\sigma^2}+\frac{1/4+2\eta-3\nu}{\sigma^2}; \quad n=0,1,2,\dots,\lambda-1
\label{EnerPTpot}
\end{eqnarray}
which corresponds to the case in which the double heterostructure ceases to be so.
\subsection{Complement of the DH defined in Table \ref{TFigT5}.} \label{SymDist2}
\begin{figure}\centering
\includegraphics[scale=0.5]{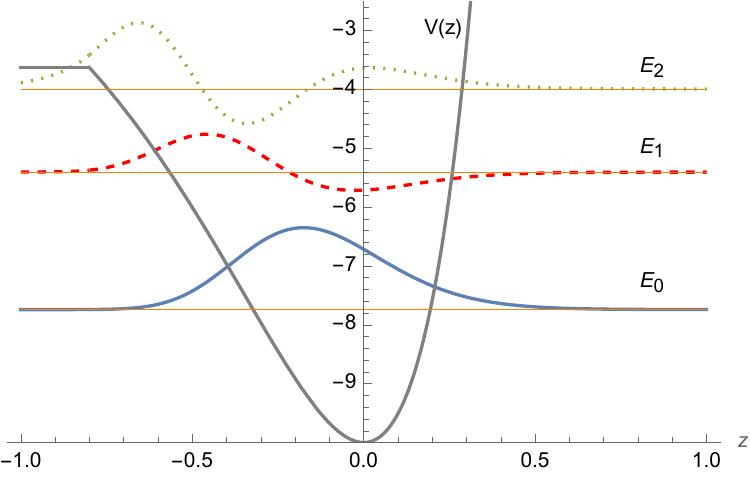}
\caption{Wave function \eqref{gensol} with $\psi_{\text{in}}^{1}(z)$, $\psi_{\text{in}}^{1}(z)$ given by Eqs. \eqref{FWitt1}.
The energy values $\{E_{0},E_{1},E_{2}\}$ are found in Table \ref{TFigT5}.}
\label{FigFig4}
\end{figure}

Let's examine a second system with analytical solutions, based on the distributions of the DH defined in Table \ref{TFigT5}
\begin{eqnarray}
V_{\textrm{in}}(z)=-V_{0}^Me^{\sigma z}(2-e^{\sigma z}), \quad 
m_{\textrm{in}}(z)=m_{0}^M \sigma^2e^{-2\sigma z},
\label{MorsPot}
\end{eqnarray}
The associated Schrödinger equation \eqref{SchomassC} is
\begin{eqnarray}
\frac{d^2\phi}{d\rho^2}+  \left(\kappa^2-\frac{2V_{0}^M\sqrt{m_{0}^M}}{\rho} - \frac{3/4+m_{0}^M V_{0}^M+2\eta-2\nu}{\rho^2}\right)\phi=0;
\quad
\rho=-\sqrt{m_0^M}e^{-\sigma z},
\label{MorseSch}
\end{eqnarray}
$\kappa^2=E$. For the case that concerns us, viz. bound states, $E=-|E|$. Eq. \eqref{MorseSch} can be rewritten as
\begin{eqnarray}
\frac{d^2\phi}{dy^2} +  \left(-\frac{1}{4}+\frac{q}{y}+\frac{1/4-r^2}{y^2}\right) \phi = 0,
\label{MorseSch2}
\end{eqnarray}
where $y=-2i\kappa \rho$, $q=-V_{0}^M \sqrt{m_{0}^M}/\sqrt{|E|}$, $r^2=1+m_{0}^MV_{0}^M+2\eta - 2\nu$. Taken into account the solutions \cite{abramowitz1965}, $M_{q,r}(y), W_{q,r}(y)$, of \eqref{MorseSch2}, the exact solutions \eqref{transf} for this system \eqref{MorsPot} are 
\begin{align}
\psi_{\text{in}}^{1}(z) = \left(m_0^{M}\right)^{1/4} \sqrt{\sigma} \, e^{-\sigma z / 2} M_{q,r}(y);
\quad
\psi_{\text{in}}^{2}(z) = \left(m_0^{M}\right)^{1/4} \sqrt{\sigma} \, e^{-\sigma z / 2} W_{q,r}(y). \label{FWitt1}
\end{align}
$M_{q,r}(y)$ and $W_{q,r}(y)$ are the Whittaker functions of first and second kind respectively. Figure \eqref{FigFig4} illustrates three functions along with their energies calculated from Eq. \eqref{trasEq}. Table \eqref{TFigT5} compares these results with values derived from the coefficient poles.

\section{}\label{AppendB}
Another potential solution to Equation \eqref{shoeqn} is by substituting \(\psi(z) = m(z)^{1/2} \Phi\). This substitution leads to the following equation
\begin{eqnarray}
\left(-\frac{d^2}{dz^2}+V^{*}(z)-E^{*}\right) \Phi= 0.
\label{SchomassC2}
\end{eqnarray}
where
\begin{eqnarray}
V^{*}(z)-E^{*}:=
\frac{1}{4}\frac{m'(z)^2}{m(z)^2}(3+2\eta)-\frac{1}{2}\frac{m''(z)}{m(z)}(1+\nu) + (V(z) - E)m(z).
\end{eqnarray}
Eq. \eqref{SchomassC2} can be solved as long as $V^{*}(z)$ is an exact, quasi-exact, or conditionally exact potential, and $E^{*}$ is a constant. 

\subsection{Application to the DH defined in Table \ref{TFigT8}}\label{b1}
For the profiles of DH of Section \ref{SExpMass}
\begin{eqnarray*}
V_{\textrm{in}}(z)=V_{c}e^{cz}; \quad 
m_{\textrm{in}}(z)=\mu_{0}e^{cz}; \quad \mu_{0}, V_{c}>0.
\end{eqnarray*}
we get the equation
\begin{eqnarray}
\left(-\frac{d^2}{dz^2}+2\mu_{0}e^{cz}(V_{c}e^{cz}-E)-E^{*}\right) \Phi=0; \quad
E^{*}=\frac{c^2(-1-2 \eta+2\nu)}{4}.
\label{EqMorseV}
\end{eqnarray}
Or
\begin{eqnarray}
\left(-\frac{d^2}{dx^2}+De^{cx}(e^{cx}-2)-E^{*}\right) \Phi=0; \quad
D=\frac{E^2\mu_{0}}{V_{c}}, \quad e^{cx}=\frac{2V_{c}}{E}e^{cz}.
\label{TipoMorse2}
\end{eqnarray}
Following the standard procedure, the ansatz $\Phi=y^s e^{-y/2}F(y)$ let us to confluent hypergeometric equation
\begin{eqnarray}
y \frac{d^2F}{dy^2}+(b-y)\frac{dF}{dy}-aF=0; \quad a=\frac{1}{2}+s-\lambda, \quad b=1+2s,
\label{HypGEqK}
\end{eqnarray}
where has been defined $y=2\lambda e^{cx}$, $\lambda=\sqrt{D}/c$, $s^2=-E^{*}/c^2$, and whose solutions are the Kummer's function of the first kind $M(a,b;y)={}_{1}F_{1}(a,b;y)$ and the Tricomi confluent hypergeometric function 
\begin{eqnarray}
U(a,b;y)=\frac{\Gamma(1-b)}{\Gamma(a+1-b)}M(a,b;y)+\frac{\Gamma(b-1)}{\Gamma(a)}z^{1-b}M(a+1-b,2-b;y).
\label{UMSol}
\end{eqnarray}

\subsection{Application to the DH defined in Table \ref{FigT9}}\label{b2}
On the other hand, if we used the potencial and mass given by \eqref{singmass}
$$
m_{\textrm{in}}(z)=c z^2;\quad
V_{\textrm{in}}(z)=\frac{A}{c} \left(\frac{1}{z^4}+\frac{B}{z^2}\right), \quad z,c,A,-B>0.
$$ 
we obtain the Schrödinger equation
\begin{eqnarray}
-\frac{d^2}{dz^2}\Phi(z)+ \left( \frac{1}{4}\omega^2z^2+\frac{g}{2z^2}-E^{*}\right)\Phi(z)=0,
\end{eqnarray}
whose energy spectra is \cite{Calogero,DZhu_1987,Cariñena_2008,Ikhdair,Sesma_2010}
\begin{eqnarray}
E^{*}=\omega \left(2n+1+\frac{1}{2}\sqrt{1+2g}\right).
\label{enerIsOs}
\end{eqnarray}
For this example $E^{*}=-AB$, $\omega^2=-4cE$, and $g=2(2+2\eta-\nu)+2A$.
\end{appendices} 

\bibliographystyle{unsrt}
\bibliography{paperCF}

\newpage
\section{Tables}
\begin{table}[h] \centering 
\begin{tabular}{@{}lccccccc@{}} 
\toprule
\multicolumn{6}{c}{Symmetric Double Heterostructure \eqref{modelmass}} &\multirow{2}{*}{\includegraphics[scale=0.2]{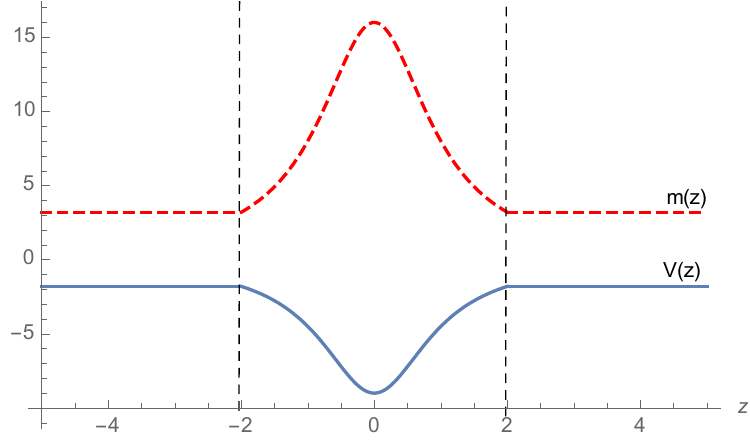}}\\ 
\multicolumn{6}{c}{$V_{\textrm{in}}(z)=-\mu^2/(1+z^2), \quad m_{\textrm{in}}(z)=\sigma^2/(1+z^2)$} \\ 
\multicolumn{6}{c}{$V_{0}=V_{2}=V_{\textrm{in}}(z_{0})$, $m_{0}=m_{2}=m_{\textrm{in}}(z_{0})$.}\\ 
\multicolumn{5}{c}{}&\multicolumn{3}{c}{$\sigma=4$, $\mu=3$, $z_{0}=-z_{1}=-2$.}\\ \midrule
\multicolumn{8}{c}{BD-D KEO $(\alpha=\gamma=0)$; boundary condition \eqref{bcondvonR}.}  \\ \midrule
                                         &$E_{0}$     &$E_{1}$    &{$E_{2}$}     & $E_{3}$     & $E_{4}$     & $E_{5}$     & $E_{6}$     \\ 
Transcendental Eq. \eqref{trasEq}       & -8.25    & -6.875       & -5.625       & -4.50009    & -3.5013     & -2.63724    & -1.96428 \\
Eq. \eqref{refcoefpol}, $R_{c}$ poles   & -8.25    & -6.875       & -5.625       & -4.50009    & -3.5013     & -2.63724    & -1.96428 \\
Eq. \eqref{EnerPTpot}, $\mathbf{E}_{n}$ & -8.25    & -6.875       & -5.625       & -4.5        & -3.5        & -2.625      & -1.875   \\ \midrule

\multicolumn{8}{c}{Z-K KEO $(\alpha=\gamma=-1/2)$; boundary condition \eqref{bcondvonR}.} \\ \midrule
Transcendental Eq. \eqref{trasEq}       & -8.3099   & -6.9297      & -5.6745      & -4.54428    & -3.53899    & -2.66042    & -1.94466 \\
Eq. \eqref{refcoefpol}, $R_{c}$ poles   & -8.3099   & -6.9297      & -5.6745      & -4.54428    & -3.53899    & -2.66042    & -1.94466 \\
Eq. \eqref{EnerPTpot}, $\mathbf{E}_{n}$ & -8.3099    & -6.9297      & -5.6745      & -4.54430    & -3.539103   & -2.65890    & -1.90370 \\ \midrule

\multicolumn{8}{c}{$T_{L}$ KEO $(\alpha \neq\gamma)$; boundary condition \eqref{bcondNew}.} \\ \midrule
\rowcolor[HTML]{B8B7B7}
Eq. \eqref{refcoefpol}, $R_{c}$ poles  & -8.29051   & -6.9132     &-5.66088       & -4.53358    & -3.53155    & -2.65848    & -1.95561 \\
\rowcolor[HTML]{B8B7B7}
Eq. \eqref{EnerPTpot}, $\mathbf{E}_{n}$& -8.29167   & -6.91667    &-5.66667       & -4.54167    & -3.54167    & -2.66667    & -1.91667 \\  \bottomrule
\end{tabular}
\caption{The energy spectrum of the DH depends on the KEO. 
Practically, the energy spectrum of the case $\alpha \neq \gamma$  is found to lie between the spectra of the cases $\alpha=\gamma$. 
The energy spectrum, $\mathbf{E}_{n}$ \eqref{EnerPTpot}, has been included as reference, it is obtained when $|z_{0}|\rightarrow \infty$. 
The boundary conditions \eqref{bcondvonR} (\eqref{bcondNew}) were used in the calculation of the reflection coefficient poles for the BD-D, Z-K  and L KEOs in all tables.
For the next references (tables), we define $V_{s}(z)=-\mu^2/(1+z^2)$ and $m_{s}(z)=\sigma^2/(1+z^2)$. 
The highlighted lines are contributions of this article.}
\label{tablesys1}
\end{table}

\begin{table}[h] \centering
\begin{tabular}{lccccccc}
\hline
\multicolumn{6}{c}{Double Heterostructure \eqref{modelmass}} &\multirow{2}{*}{\includegraphics[scale=0.2]{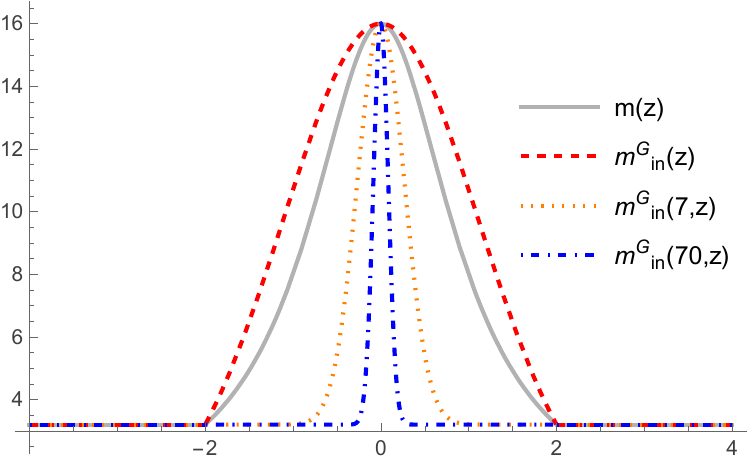}} \\
\multicolumn{6}{c}{$V_{\textrm{in}}(z)=V_{s}(z),\quad m_{\textrm{in}}(z)=m_{\textrm{in}}^{G}(z)\quad \text{or} \quad m_{\textrm{in}}^{G}(\delta,z).$}\\
\multicolumn{6}{c}{$V_{0}=V_{2}=V_{\textrm{in}}(z_{0})$, $m_{0}=m_{2}=m_{\textrm{in}}(z_{0})$.}\\ 
\multicolumn{5}{c}{}& \multicolumn{3}{c}{$\sigma=4$, $\mu=3$, $z_{0}=-z_{1}=-2.$}                                                 \\ \hline
\multicolumn{8}{c}{BD-D KEO $(\alpha=\gamma=0)$; boundary condition \eqref{bcondvonR}.}                         \\ \hline
           & $E_0$       & $E_1$       & $E_2$       & $E_3$       & $E_4$       & $E_5$       & $E_6$         \\
$m_{\textrm{in}}^{G}(z)$       & -8.27528 & -6.92737 & -5.727   & -4.66314 & -3.72403 & -2.9007  & -2.20866   \\
$m_{s}(z)$           & -8.25    & -6.875   & -5.625   & -4.50009 & -3.5013  & -2.63724 & -1.96428   \\
$m_{\textrm{in}}^{G}(7,z)$     & -8.01692 & -6.47117 & -5.00746 & -3.52209 & -2.28029 &          &            \\
$m_{\textrm{in}}^{G}(70,z)$    & -7.57495 & -6.13984 & -3.45826 & -2.56635 &          &          &            \\ \hline
\multicolumn{8}{c}{Z-K KEO $(\alpha=\gamma=-1/2)$; boundary condition \eqref{bcondvonR}.}                      \\ \hline
$m_{\textrm{in}}^{G}(z)$       & -8.30142 & -6.95596 & -5.75918 & -4.70081 & -3.76994 & -2.95444 & -2.24512   \\
$m_{s}(z)$        & -8.3099  & -6.9297  & -5.6745  & -4.54428 & -3.53899 & -2.66042 & -1.94466   \\
$m_{\textrm{in}}^{G}(7,z)$  & -8.37197 & -6.63962 & -4.83016 & -3.36362 & -2.32209 &          &            \\
$m_{\textrm{in}}^{G}(70,z)$ & -8.26032 & -5.19903 & -4.0049  & -2.2545  &          &          &            \\ \hline
\multicolumn{8}{c}{$T_{L}$ KEO $(\alpha \neq\gamma)$; boundary condition \eqref{bcondNew}.}                    \\ \hline
\rowcolor[HTML]{C0C0C0}
$m_{\textrm{in}}^{G}(z)$       & -8.29282 & -6.94682 & -5.74927 & -4.68975 & -3.7573  & -2.94111 & -2.23907   \\
$m_{s}(z)$        & -8.29051 & -6.9132  & -5.66088 & -4.53358 & -3.53155 & -2.65848 & -1.95561   \\
\rowcolor[HTML]{C0C0C0}
$m_{\textrm{in}}^{G}(7,z)$  & -8.27014 & -6.6162  & -4.9169  & -3.43401 & -2.31913 &          &            \\
\rowcolor[HTML]{C0C0C0}
$m_{\textrm{in}}^{G}(70,z)$ & -8.04052 & -5.53761 & -3.87688 & -2.40156 &          &          &             \\ \hline
\end{tabular}
\caption{Numerical energy spectra calculated from the poles of the reflection coefficient \eqref{refcoefpol} of the distributions
$m_{\textrm{in}}^{G}(z)=\sigma^2 e^{-(z^2/z_{0}^2)\ln{(1+z_{0}^2)}}$, $m_{\textrm{in}}^{G}(\delta,z)=\frac{\sigma^2}{1+z_{0}^2} \left(z_{0}^2e^{-\delta z^2}+1\right); \quad \delta>0$. 
The energy spectrum of the DH depends on the KEO. 
Note that the DH in Table \ref{tablesys1} is being considered in $m_{s}(z)$. 
The highlighted lines are contributions of this article.}
\label{TSymmGauss1}
\end{table}

\begin{table}[h] \centering
\begin{tabular}{lcccccccc}
\hline
\multicolumn{7}{c}{Double Heterostructure \eqref{modelmass}} &\multirow{2}{*}{\includegraphics[scale=0.2]{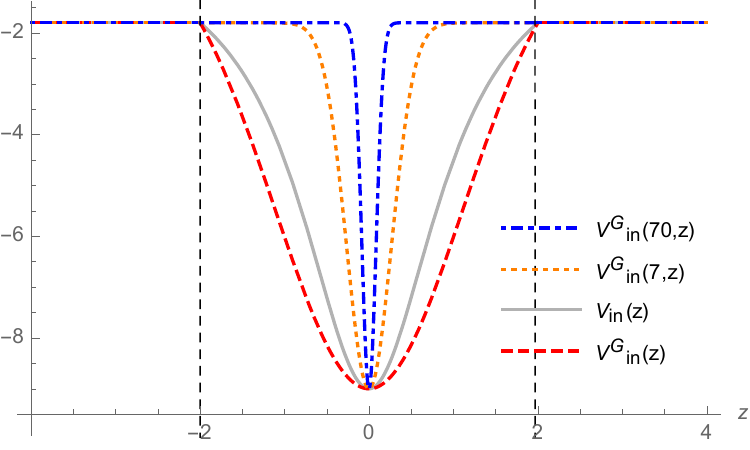}} \\
\multicolumn{7}{c}{$m_{\textrm{in}}(z)=m_{s}(z),\quad V_{\textrm{in}}(z)=V_{\textrm{in}}^{G}(z) \quad \text{or} \quad V_{\textrm{in}}^{G}(\delta,z).$}\\
\multicolumn{7}{c}{$V_{0}=V_{2}=V_{\textrm{in}}(z_{0})$, $m_{0}=m_{2}=m_{\textrm{in}}(z_{0})$.}\\ 
\multicolumn{6}{c}{} & \multicolumn{3}{c}{$\sigma=4$, $\mu=3$, $z_{0}=-z_{1}=-2.$}                    \\ \hline
\multicolumn{9}{c}{BD-D KEO $(\alpha=\gamma=0)$; boundary condition \eqref{bcondvonR}.}                                                         \\ \hline
                            & $E_0$       & $E_1$       & $E_2$      & $E_3$     & $E_4$     & $E_5$       & $E_6$       & $E_{7}$  \\
$V_{\textrm{in}}^{G}(z)$    & -8.48885    & -7.52102    & -6.5416    & -5.55965  & -4.58509  & -3.63137    & -2.72542    & -1.95856 \\
$V_{\textrm{in}}^{G}(1,z)$  & -8.30802    & -7.00905    & -5.76342   & -4.59031  & -3.51713  & -2.59422    & -1.97481    &  \\ 
$V_{s}(z)$                  & -8.25       & -6.875      & -5.625     & -4.50009  & -3.5013   & -2.63724    & -1.96428    &  \\
$V_{\textrm{in}}^{G}(7,z)$  & -7.34722    & -4.45967    & -2.41175   &           &           &             &             &  \\ \hline
\multicolumn{9}{c}{Z-K KEO $(\alpha=\gamma=-1/2)$; boundary condition \eqref{bcondvonR}.}                                   \\ \hline
$V_{\textrm{in}}^{G}(z)$    & -8.54777   & -7.57359    & -6.58914    & -5.60293  & -4.62408  & -3.66359 & -2.74035 & -1.93135  \\
$V_{\textrm{in}}^{G}(1,z)$  & -8.36777   & -7.06359    & -5.81314    & -4.63529  & -3.55622  & -2.61807 & -1.95067 &  \\
$V_{s}(z)$                  & -8.3099    & -6.9297     & -5.6745     & -4.54428  & -3.53899  & -2.66042 & -1.94466 &  \\
$V_{\textrm{in}}^{G}(7,z)$  & -7.40843   & -4.51762    & -2.46253    &           &           &          &          &  \\ \hline
\multicolumn{9}{c}{$T_{L}$ KEO $(\alpha \neq\gamma)$; boundary condition \eqref{bcondNew}.}                                                         \\ \hline
\rowcolor[HTML]{C0C0C0}
$V_{\textrm{in}}^{G}(z)$    &-8.52892    &-7.55827    &-6.57661  &-5.59274  &-4.61611  &-3.65847  &-2.74106  &-1.94436  \\
\rowcolor[HTML]{C0C0C0}
$V_{\textrm{in}}^{G}(1,z)$  &-8.34846    &-7.04718    &-5.7994   &-4.62416  &-3.5481   &-2.61575  &-1.9613   &  \\ 
$V_{s}(z)$                  & -8.29051   & -6.9132    & -5.66088 & -4.53358 & -3.53155 & -2.65848 & -1.95561 &  \\
\rowcolor[HTML]{C0C0C0}
$V_{\textrm{in}}^{G}(7,z)$  &-7.38831    &-4.4993   &-2.4482   &         &          &          &           & \\ \hline
\end{tabular}
\caption{Numerical energy spectra calculated from the poles of the reflection coefficient \eqref{refcoefpol} of the distributions
$V_{\textrm{in}}^{G}(z)=-\mu^2e^{-(z^2/z_{0}^2)\ln{(1+z_{0}^2)}}$, 
$V_{\textrm{in}}^{G}(\delta,z)=-\frac{\mu^2}{1+z_{0}^2} \left(z_{0}^2e^{-\delta z^2}+1\right); \quad \delta>0$. Note that the DH in Table \ref{tablesys1} is being considered in $V_{s}(z)$. The highlighted lines are contributions of this article.}
\label{TSymmGauss2}
\end{table}

\begin{table}[h] \centering
\begin{tabular}{@{}lccc@{}}
\toprule
\multicolumn{3}{c}{Double Heterostructure \eqref{modelmass}}  &\multirow{2}{*}{\includegraphics[scale=0.2]{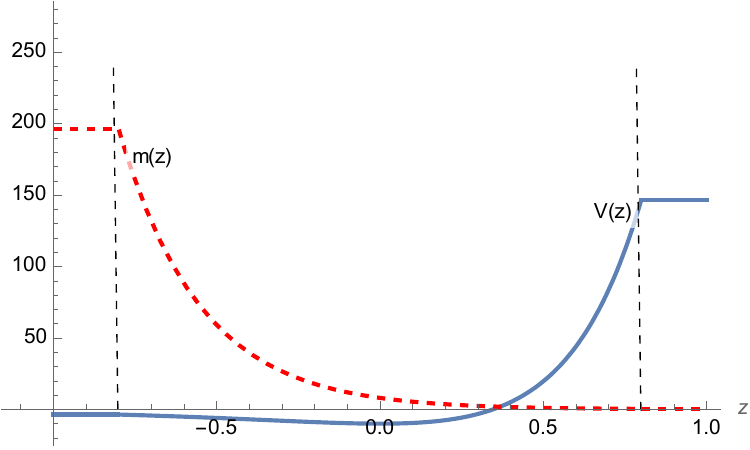}} \\
\multicolumn{3}{c}{$V_{\textrm{in}}(z)=-V_{0}^Me^{\sigma z}(2-e^{\sigma z})$, $m_{\textrm{in}}(z)=m_{0}^M \sigma^2e^{-2\sigma z}$}\\
\multicolumn{3}{c}{$V_{0,2}=V_{\textrm{in}}(z_{0,1})$, $m_{0,2}=m_{\textrm{in}}(z_{0,1})$.} \\
\multicolumn{3}{c}{}&{$\sigma=2$, $z_{0}=-z_{1}=-0.8$} \\ \midrule
\multicolumn{4}{c}{BD-D KEO $(\alpha=\gamma=0)$; boundary condition \eqref{bcondvonR}.}                                                                \\ \midrule
                           &$E_{0}$   &$E_{1}$  &$E_{2}$    \\ 
Transcendental Eq. \eqref{trasEq},      & -7.74229 & -5.40587   & -3.99419                    \\
\multicolumn{1}{l}{Poles}               & -7.74229 & -5.40587   & -3.99419                    \\ \midrule
\multicolumn{4}{c}{Z-K KEO $(\alpha=\gamma=-1/2)$; boundary condition \eqref{bcondvonR}.}                                                                 \\ \midrule 
Transcendental Eq. \eqref{trasEq},      & -8.08993 & -5.60758   & -4.12556                    \\
\multicolumn{1}{l}{Poles} & -8.08993 & -5.60758   & -4.12556                    \\ \midrule
\multicolumn{4}{c}{$T_{L}$ KEO $(\alpha \neq\gamma)$; boundary condition \eqref{bcondNew}.}                                                                \\ \midrule
\rowcolor[HTML]{B8B7B7}
\multicolumn{1}{l}{Poles} & -8.04977 & -5.5844  & -4.10875 \\ \bottomrule
\end{tabular}
\caption{Energy spectra of the DH with profiles defined by Eq. \eqref{MorsPot}. $V_{0}^M$, $m_{0}^M$, $\sigma>0$. The analytical and numerical results agree. For the next references (tables), we define $V_{M}(z)=-V_{0}^Me^{\sigma z}(2-e^{\sigma z})$ and $m_{M}(z)=m_{0}^M \sigma^2e^{-2\sigma z}$. The highlighted lines are contributions of this article; $m_{0}^{M}=2$, $V_{0}^M=10$.}
\label{TFigT5}
\end{table}

\begin{table}[h] \centering
\begin{tabular}{@{}lcccccc@{}}
\hline
\multicolumn{5}{c}{Double Heterostructure \eqref{modelmass}} &\multirow{2}{*}{\includegraphics[scale=0.2]{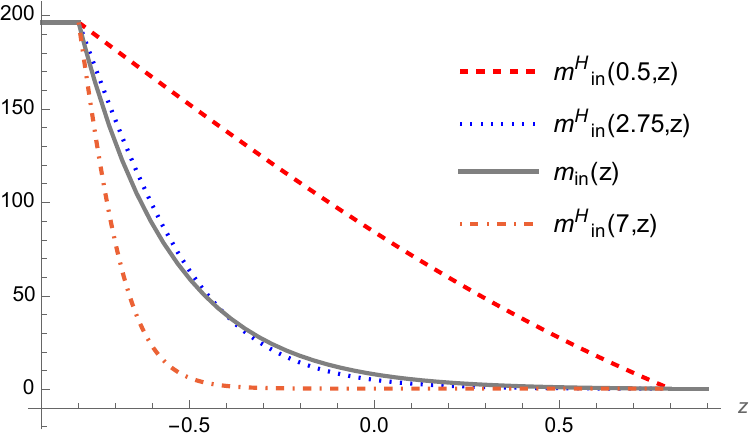}}     \\
\multicolumn{5}{c}{$V_{\textrm{in}}(z)=V_{M}(z), \quad m_{\textrm{in}}(z)=m_{\textrm{in}}^H(\tau,z)$}\\
\multicolumn{5}{c}{$V_{i}=V_{\textrm{in}}(z_{i})$, $m_{i}=m_{\textrm{in}}(z_{i})$, $i=0,2$.} \\
\multicolumn{4}{c}{} &\multicolumn{3}{c}{$\sigma=2$, $z_{0}=-z_{1}=-0.8$.} \\ \hline
\multicolumn{7}{c}{BD-D KEO $(\alpha=\gamma=0)$; boundary condition \eqref{bcondvonR}.}  \\ \hline
                            & $E_{0}$          & $E_{1}$          & $E_{2}$      & $E_{3}$       & $E_{4}$        & $E_{5}$       \\
$m_{\textrm{in}}^H(0.5,z)$  & -8.07565         & -6.96429         & -5.9702      & -5.08382      & -4.29792        & -3.65763      \\
$m_{M}(z)$        & -7.74229         & -5.40587         & -3.99419     &               &                &               \\

$m_{\textrm{in}}^H(2.75,z)$ & -7.15447        & -4.99484         & -3.7632       &                &                &              \\
$m_{\textrm{in}}^H(7,z)$    & -3.69866        &                  &               &                &                &              \\ \hline
\multicolumn{7}{c}{Z-K KEO $(\alpha=\gamma=-1/2)$; boundary condition \eqref{bcondvonR}.} \\ \hline
$m_{\textrm{in}}^H(0.5,z)$  & -8.09038         & -6.97779        & -5.98252       & -5.09502        & -4.30839        & -3.66699     \\
$m_{M}(z)$        & -8.08993         & -5.60758        & -4.12556      &                &                &              \\
$m_{\textrm{in}}^H(2.75,z)$ & -7.76532         & -5.29044        & -3.93178       &                &                &              \\
$m_{\textrm{in}}^H(7,z)$    & -4.21368         &                 &               &                &                &              \\ \hline
\multicolumn{7}{c}{$T_{L}$ KEO $(\alpha \neq\gamma)$; boundary condition \eqref{bcondNew}.} \\ \hline
\rowcolor[HTML]{C0C0C0}
$m_{\textrm{in}}^H(0.5,z)$  & -8.08676         & -6.977447       & -5.97948      & -5.09226       & -4.30576       & -3.66431     \\ 
$m_{M}(z)$        & -8.04977         & -5.5844         & -4.10875      &                &                &              \\
\rowcolor[HTML]{C0C0C0}
$m_{\textrm{in}}^H(2.75,z)$ & -7.70847         & -5.2647         & -3.9123       &                &                 &             \\
\rowcolor[HTML]{C0C0C0}
$m_{\textrm{in}}^H(7,z)$    & -4.4308          &                 &               &                &                 &             \\ \hline        
\end{tabular}
\caption{Energy spectrum of the DH system with hyperbolic mass $m_{\textrm{in}}^H(\tau,z)=m_{0}^M \sigma^2 \left((e^{-2\sigma z_{1}}-e^{-2\sigma z_{0}})B(z)+e^{-\sigma z_{0}}\right); \quad 
B(z)=\frac{\tanh{[-\tau(z-z_{0})}]}{\tanh{[-\tau(z_{1}-z_{0})}]}$. 
Note that the DH in Table \ref{TFigT5} is being considered in $m_{M}(z)$.
The highlighted lines are contributions of this article;
$m_{0}^M=2$, $V_{0}^M=10$.}
\label{TFigT6}
\end{table}

\begin{table}[h]\centering
\begin{tabular}{lccc}
\hline
\multicolumn{3}{c}{Double heterostructure \eqref{modelmass}} &\multirow{2}{*}{\includegraphics[scale=0.2]{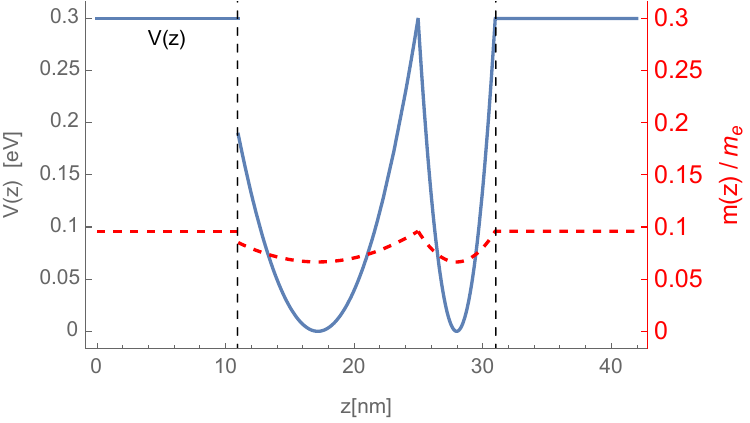}}  \\ 
\multicolumn{3}{c}{$V_{\textrm{in}}(z)=V_{p}(z)$, $m_{\textrm{in}}(z)=m_{p}(z)$ \eqref{parabPot}} \\
\multicolumn{3}{c}{$V_{0}=V_{2}=0.3$ eV, $m_{1}=0.0655 m_{e}$, $m_{0}=0.0960 m_{e}$.}                          \\ \hline
                    & $E_{0}$        & $E_{1}$          & $E_{2}$          \\ \hline
$R_{c}^{BD-D}$ Poles    & 50.5284   & 117.34107   & 156.51738   \\
$R_{c}^{Z-K}$  Poles    & 54.1197   & 130.65338   & 160.38424   \\
\rowcolor[HTML]{C0C0C0}
$R_{c}^{L}$ Poles    & 52.8956   & 126.1985    & 158.91771   \\ \hline
\end{tabular}
\caption{Bound state energies (meV) of an asymmetrical parabolic quantum well pair with position-dependent mass \eqref{parabPot}; $a=9.4$ nm, $b=11$ nm, $c=25$ nm, $d=31$ nm.
The highlighted lines are contributions of this article.}
\label{TPotParab}
\end{table}

\begin{table}[h] \centering
\begin{tabular}{@{}lcccc@{}}
\toprule
\multicolumn{4}{c}{Double Heterostructure \eqref{modelmass}}  &\multirow{2}{*}{\includegraphics[scale=0.2]{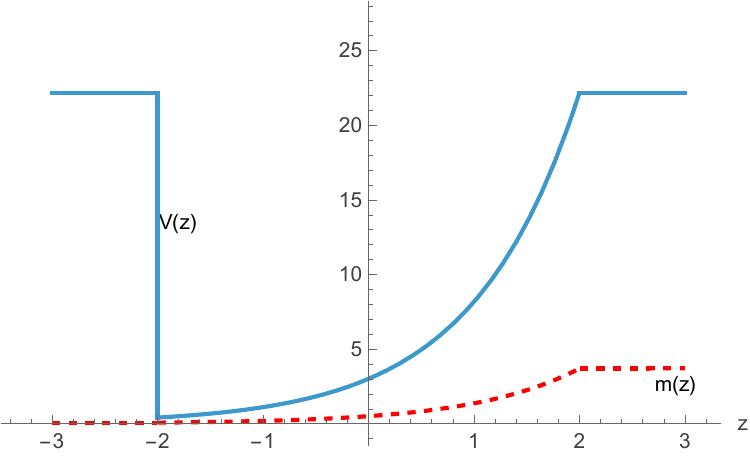}} \\
\multicolumn{4}{c}{$V(z)$ and $m(z)$ given by Eq.  \eqref{FigT8}.} \\
\multicolumn{4}{c}{$V_{c}=3$, $\mu_{0}=1/2$, $\lambda=1$.}\\
\multicolumn{4}{c}{}& $c=1$, $z_{0}=-2$, $z_{1}=2$ \\ \midrule
\multicolumn{5}{c}{BD-D KEO $(\alpha=\gamma=0)$; boundary condition \eqref{bcondvonR}.}                                                \\ \midrule
                                                            &$E_{0}$    &$E_{1}$     &$E_{2}$   &$E_{3}$    \\ 
Transcendental Eq. \eqref{trasEq},                          & 5.13516   &10.1865     &15.1575   &19.9185  \\
\multicolumn{1}{l}{Eq. \eqref{refcoefpol}, $R_{c}$ poles}   & 5.13456   &10.1856     &15.1565   &19.9177  \\ \midrule
\multicolumn{5}{c}{Z-K KEO $(\alpha=\gamma=-1/2)$; boundary condition \eqref{bcondvonR}.}                                                \\ \midrule 
Transcendental Eq. \eqref{trasEq},                          &4.63268  &10.0389    &15.223       &20.076    \\
\multicolumn{1}{l}{Eq. \eqref{refcoefpol}, $R_{c}$ poles}   &4.6323   &10.0384    &15.2222      &20.0753    \\ \midrule
\multicolumn{5}{c}{$T_{L}$ KEO $(\alpha \neq\gamma)$; boundary condition \eqref{bcondNew}.}                                                                \\ \midrule
\multicolumn{1}{l}{Eq. \eqref{refcoefpol}, $R_{c}$ poles}   &4.63027 &9.9751  &15.119 &19.965 \\ \bottomrule
\end{tabular}
\caption{Energy spectra of the DH \eqref{FigT8} for various ambiguity parameters. 
The first column denotes the method employed for the calculation.
Note that this DH allows the BD-D spectrum, which was considered forbidden in \cite{DESOUZADUTRA200025} for GH.}
\label{TFigT8}
\end{table}

\begin{table}[h] \centering
\begin{tabular}{@{}lccc@{}}
\toprule
\multicolumn{3}{c}{Double Heterostructure \eqref{modelmass}}  &\multirow{2}{*}{\includegraphics[scale=0.2]{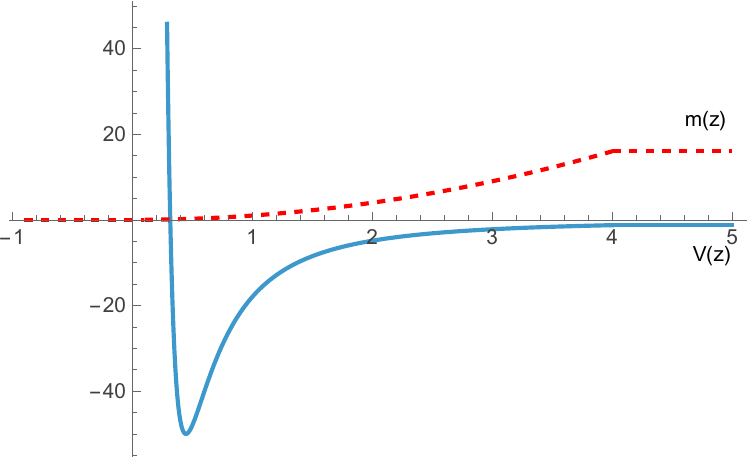}} \\
\multicolumn{3}{c}{$V(z)$ and $m(z)$ given by Eq. \eqref{singmass}}\\
\multicolumn{3}{c}{$V_{0,2}=V_{\textrm{in}}(z_{0,1})$, $m_{0,2}=m_{\textrm{in}}(z_{0,1})$.} \\
\multicolumn{3}{c}{} &{$c=1$, $z_{0}=0.1$, $z_{1}=4$.} \\ \midrule
\multicolumn{4}{c}{BD-D KEO $(\alpha=\gamma=0)$; boundary condition \eqref{bcondvonR}.}                                                 \\ \midrule
                                                            &$E_{0}$    &$E_{1}$     &$E_{2}$    \\ 
Transcendental Eq. \eqref{trasEq},                          & -10.68215 &-3.90650    &-2.00662   \\
\multicolumn{1}{l}{Eq. \eqref{refcoefpol}, $R_{c}$ poles}   & -10.6822  &-3.90651    &-2.00662   \\
\multicolumn{1}{l}{Eq. \eqref{EnSinPot}, ${\mathbf{E}_{n}}^{BD-D}$}  & -10.6688  &-3.9033     &-2.00539   \\ \midrule
\multicolumn{4}{c}{Z-K KEO $(\alpha=\gamma=-1/2)$; boundary condition \eqref{bcondvonR}.}                                                \\ \midrule 
Transcendental Eq. \eqref{trasEq},                          &-16.05884  & -4.94871   &-2.370368    \\
\multicolumn{1}{l}{Eq. \eqref{refcoefpol}, $R_{c}$ poles}   &-16.05698  &-4.94871    &-2.37037     \\ 
\multicolumn{1}{l}{Eq. \eqref{EnSinPot}, ${\mathbf{E}_{n}}^{Z-K}$}  &-16.       &-4.93827    &-2.36686    \\\midrule
\multicolumn{4}{c}{$T_{L}$ KEO $(\alpha \neq\gamma)$; boundary condition \eqref{bcondNew}.}                                                                \\ \midrule
\multicolumn{1}{l}{Eq. \eqref{refcoefpol}, $R_{c}$ poles}   &-14.38634 &-4.65256  &-2.27083 \\ \bottomrule
\end{tabular}
\caption{Energy spectra of the DH \eqref{singmass} for various ambiguity parameters; 
$A=2$, $B=-10$.}
\label{FigT9}
\end{table}

\end{document}